\documentclass[floats,floatfix,showpacs,amssymb,prd,twocolumn,superscriptaddress,nofootinbib]{revtex4-1}
\usepackage[dvips]{graphicx}
\usepackage{amsmath}
\usepackage{url}
\usepackage{color}
\usepackage{graphicx}  % needed for figures
\usepackage{bm}        % for math
\usepackage{amssymb}   % for math
\usepackage{hyperref}
\definecolor{linkcolor}{rgb}{0.0,0.3,0.5}
\definecolor{urlcolor}{rgb}{0.27,0.55,0.}
\definecolor{funcolor}{rgb}{0.65, 0.16, 0.16}
\hypersetup{colorlinks=true,linkcolor=linkcolor,citecolor=linkcolor,filecolor=linkcolor,urlcolor=linkcolor}

\begin{document}
	
\title{Nutational resonances, transitional precession, and precession-averaged evolution in binary black-hole systems}
	
\author{Xinyu Zhao}
\email{xxz132230@utdallas.edu}

\author{Michael Kesden}
\email{kesden@utdallas.edu}
\affiliation{Department of Physics, The University of Texas at Dallas, Richardson, Texas 75080, USA}
	
\author{Davide Gerosa}
\thanks{Einstein Fellow}
\email{dgerosa@caltech.edu}
\affiliation{TAPIR 350-17, California Institute of Technology, 1200 E. California Blvd., Pasadena, California 91125, USA}
	
\date{\today}
	
\begin{abstract}
In the post-Newtonian (PN) regime, the timescale on which the spins of binary black holes precess is much shorter than the
radiation-reaction timescale on which the black holes inspiral to smaller separations.  On the precession timescale, the angle
between the total and orbital angular momenta oscillates with nutation period $\tau$, during which the orbital angular
momentum precesses about the total angular momentum by an angle $\alpha$.  This defines two distinct frequencies that
vary on the radiation-reaction timescale: the nutation frequency $\omega \equiv 2\pi/\tau$ and the precession frequency
$\Omega \equiv \alpha/\tau$.  We use analytic solutions for generic spin precession at 2PN order to derive Fourier series for the
total and orbital angular momenta in which each term is a sinusoid with frequency $\Omega - n\omega$ for integer $n$.  As
black holes inspiral, they can pass through nutational resonances ($\Omega = n\omega$) at which the total angular
momentum tilts.  We derive an approximate expression for this tilt angle and show that it is usually less than $10^{-3}$
radians for nutational resonances at binary separations $r > 10M$.  The large tilts occurring during transitional precession
(near zero total angular momentum) are a consequence of such states being approximate $n=0$ nutational resonances.  Our
new Fourier series for the total and orbital angular momenta converge rapidly with $n$ providing an intuitive and
computationally efficient approach to understanding generic precession that may facilitate future calculations of
gravitational waveforms in the PN regime.
\end{abstract}
	
\maketitle
	
\section{Introduction} \label{S:intro}

The discovery of gravitational waves (GWs) emitted by binary black holes (BBHs)
\cite{2016PhRvL.116f1102A,2016PhRvL.116x1103A,2016PhRvX...6d1015A}
provides powerful motivation to better understand the generic behavior of such systems.  BBH mergers can be divided into
three stages: the inspiral during which the BBHs approach each other as their orbit decays due to radiation reaction, the
merger proper in which the two BBH event horizons coalesce into the single horizon of the final black hole, and the ringdown
in which the final black hole settles down to an unperturbed Kerr solution describing an isolated spinning black hole
\cite{1963PhRvL..11..237K}.  The evolution during each of these three stages is best described by a different numerical
technique.  The post-Newtonian (PN) approximation pioneered by Einstein himself \cite{1915SPAW.......831E} works well
during the inspiral stage when the binary separation $r$ is much greater than than the gravitational radius
$r_g \equiv GM/c^2$, where $M = m_1 + m_2$ is the sum of the BBH masses $m_i$, $G$ is Newton's gravitational constant,
and $c$ is the speed of light.  Numerical relativity
\cite{2005PhRvL..95l1101P,2006PhRvL..96k1101C,2006PhRvL..96k1102B} is required to describe the final orbits and
merger proper, while black-hole perturbation theory
\cite{1957PhRv..108.1063R,1970PhRvD...2.2141Z,1973ApJ...185..635T}
provides a good description of the late ringdown when the spacetime is close to the Kerr solution describing the final black 
hole.

This paper will focus on the inspiral stage of the merger at binary separations $r \gg r_g$ for which the PN approximation is
valid.  This stage is important for several reasons.  BBHs with $M \lesssim 25 M_\odot$ [such as the system responsible for
GW151226 \cite{2016PhRvL.116x1103A}, the second detection by the Laser Interferometer Gravitational-wave Observatory
(LIGO)] are well described by a PN inspiral when emitting GWs at the lower end of the LIGO sensitivity band.  Although the
PN regime does not fall within the LIGO band for more massive systems like the one responsible for the first LIGO detection
GW150914 \cite{2016PhRvL.116f1102A}, in the future such systems may be detectable in the PN regime at lower GW
frequencies by space-based observatories such as LISA \cite{2016PhRvL.116w1102S}.  Finally, the PN approximation
is essential for evolving BBHs from the wide separations at which they form to the smaller separations at which they emit
detectable GWs \cite{2015PhRvL.114h1103K,2015PhRvD..92f4016G,2016PhRvD..93l4066G}.  This evolution is required for
efforts to use BBH spins to distinguish between different astrophysical models of BBH formation
\cite{2010PhRvD..81h4054K,2010ApJ...715.1006K,2012PhRvD..85l4049B,2013PhRvD..87j4028G,2016ApJ...832L...2R,2017arXiv170306873S}.
	
In the PN regime, BBHs evolve on three distinct timescales:
\begin{subequations} \label{E:timescales}
\begin{align}
t_{\rm orb} &\equiv \left( \frac{r^3}{GM} \right)^{1/2}, \label{E:Torb} \\
t_{\rm pre} &\equiv \frac{c^2r^{5/2}}{(GM)^{3/2}} = \left( \frac{r}{r_g} \right) t_{\rm orb}, \label{E:Tpre} \\
t_{\rm RR} &\equiv \frac{c^5r^4}{(GM)^3} = \left( \frac{r}{r_g} \right)^{5/2} t_{\rm orb} \label{E:TRR},
\end{align}
\end{subequations}
where the direction of the binary separation vector $\mathbf{r}$ changes on the orbital timescale $t_{\rm orb}$, the directions
of the BBH spins $\mathbf{S}_i$ and orbital angular momentum $\mathbf{L}$ change on the precession timescale
$t_{\rm pre}$, and the binary separation $r$ shrinks on the radiation-reaction timescale $t_{\rm RR}$.  The validity of the PN
approximation ($r \gg r_g$) implies that these timescales obey the hierarchy \mbox{$t_{\rm orb} \ll t_{\rm pre} \ll t_{\rm RR}$}.  This
hierarchy suggests that BBH dynamics can be understood through a multi-timescale analysis: the evolution on a given 
timescale can be solved by holding constant quantities evolving on longer timescales and time-averaging quantities evolving
on shorter timescales.

In the case of BBH evolution, a multi-timescale analysis requires two different kinds of averaging: using Keplerian or higher
PN-order solutions to the two-body problem to orbit average when considering evolution on the precession or
radiation-reaction timescales, and using PN solutions to the spin-precession equations to precession average when
considering evolution on the radiation-reaction timescale.  Orbit averaging using either circular or eccentric Keplerian orbits
has been employed in many previous studies of solutions to the spin-precession equations
\cite{1994PhRvD..49.6274A,1995PhRvD..52..821K,2004PhRvD..70l4020S}.  In previous work
\cite{2015PhRvL.114h1103K,2015PhRvD..92f4016G}, we derived analytic solutions to the 2PN
spin-precession equations, allowing us to precession average BBH dynamics at this PN order for the first time.
This precession averaging has led to a vast increase in computational efficiency when evolving BBH spins on the
radiation-reaction timescale as binaries inspiral from wide separations into the LIGO band.  Readers can take advantage of
these computational savings by using the publicly available \textsc{python} module \textsc{precession}
\cite{2016PhRvD..93l4066G}.

In this paper, we make further use of precession averaging to derive a new series expansion for the generic evolution of the
orbital angular momentum $\mathbf{L}$ on the precession timescale.  This expansion is highly analogous to a Fourier series,
with amplitudes and frequencies varying on the longer radiation-reaction timescale.  This analysis is complicated by
the fact that precession exhibits two distinct frequencies.  The nutation frequency $\omega \equiv 2\pi/\tau$ is the frequency with
which the angle $\theta_L$ between $\mathbf{L}$ and the total angular momentum $\mathbf{J}$ oscillates, where $\tau$ is
the period of these oscillations.  The precession frequency $\Omega \equiv \alpha/\tau$ is the average rate at which $\mathbf{L}$
precesses in a cone about $\mathbf{J}$, where $\alpha$ is the precession angle over the nutation period $\tau$.  Each term
in our series expansion corresponds to simple precession of a vector in the plane perpendicular to the precession-averaged
$\langle \mathbf{J} \rangle$, with the magnitude of each vector fixed on the precession timescale and the precession frequency
given by $\Omega - n\omega$ for integer $n$.  The magnitude of the component of $\mathbf{L}$ parallel to
$\langle \mathbf{J} \rangle$ is chosen to maintain the proper normalization of $\mathbf{L}$.  This expansion converges rapidly with
$n$, implying that it may be useful in the construction of frequency-domain waveforms for the inspiral portion of BBH mergers.  Our
analytic solutions to the spin-precession equations have already been used for waveform construction in recent work
\cite{2017PhRvL.118e1101C,2017PhRvD..95j4004C}.  We hope that the new precession-averaged expansions for $\mathbf{L}$
and $\mathbf{J}$ developed later in this paper will be similarly useful, as variation in the direction of $\mathbf{J}$ is a major source
of error for these efforts.

Our new series expansion has also revealed the existence of nutational resonances where $\Omega = n\omega$.  At such
resonances, the precession-averaged rate $\langle d\mathbf{J}/dt \rangle$ at which the total angular momentum is radiated is
misaligned with $\langle\mathbf{J}\rangle$, implying that $\langle\mathbf{J}\rangle$ is tilting on the precession timescale.  Although
the resonance condition $\Omega = n\omega$ is finely tuned at any given binary separation $r$, generic BBHs often cross
resonances as they inspiral from wide separations towards merger.  We derive approximate expressions for the angle
$\theta_{\rm tilt}$ through which $\langle \mathbf{J} \rangle$ tilts at a resonance and show that such tilts are usually below
$10^{-3}$ radians making them negligible for the purpose of GW data analysis.  An exception is the large tilts that occur during
transition precession \cite{1994PhRvD..49.6274A}, which can be interpreted as an approximate $n=0$ nutational resonance in
much of the parameter space with near-vanishing total angular momentum ($J \simeq 0$).
	
The remainder of this paper is organized as follows. Section~\ref{S:review} reviews our previous work
\cite{2015PhRvL.114h1103K,2015PhRvD..92f4016G} on analytic solutions to the orbit-averaged spin-precession equations.
In Section~\ref{S:expansion}, we make use of these solutions to derive a new series expansion for the evolution of the
orbital angular momentum $\mathbf{L}$ on the precession timescale.  We show that only a few terms in this expansion with
the lowest values of $|n|$ are required to produce excellent agreement with full numerical solutions of the orbit-averaged
spin-precession equations, and explore the implications of this expansion for the evolution of the total angular momentum
$\mathbf{J}$.  In Section~\ref{S:resonances}, we show that $\langle \mathbf{J} \rangle$ tilts at nutational resonances
where $\Omega = n\omega$ and derive an approximate expression for the tilt angle $\theta_{\rm tilt}$ that we verify agrees
well with the tilts observed in full numerical solutions of the orbit-averaged spin-precession equations.  In Section~\ref{S:demo}, we
examine how often generic binaries encounter nutational resonances during their inspirals and the distribution of tilt angles at these
resonances.  In Sec.~\ref{S:trans}, we explore the connection between our newly discovered nutational resonances and transitional
precession near $J \simeq 0$ \cite{1994PhRvD..49.6274A}.  Some concluding remarks are provided in Section~\ref{S:disc}.  In the
rest of this paper, we will use relativists' units where $G = c = 1$.
								
\section{Review of Spin Precession} \label{S:review}	
 
Consider binary black holes on a quasicircular orbit with masses $m_1$ and $m_2$, mass ratio $q \equiv m_2/m_1\leq 1$,
total mass $M \equiv m_1+m_2$, and symmetric mass ratio $\eta \equiv m_1m_2/M^2 = q/(1+q)^2$.  Such a system will
have an orbital angular momentum $\mathbf{L}$ with magnitude $L = \eta(M^3r)^{1/2}$ to lowest PN order and spins
$\mathbf{S}_i$ with magnitudes $S_i = \chi_i m_i^2$, where the dimensionless spins have magnitudes $0 \leq \chi_i \leq 1$.
The total spin $\mathbf{S} = \mathbf{S}_1 + \mathbf{S}_2$ has magnitude $S$, and the total angular momentum
$\mathbf{J} = \mathbf{L} +\mathbf{S}$ has magnitude $J$.  Each of these quantities is either constant or evolves on one of the
timescales given by Eq.~(\ref{E:timescales}).  At the PN order we consider in this paper, the masses $m_i$ and
dimensionless spin magnitudes $\chi_i$ are constant throughout the inspiral.  The projected effective spin
\begin{equation} \label{E:ProjEff}
\xi \equiv \frac{1}{M^2} \left[\left(1+q\right)\mathbf{S}_1 + \left(1+\frac{1}{q}\right)\mathbf{S}_2\right] \cdot \hat{\mathbf{L}}~,
\end{equation}
referred to as $\chi_{\rm eff}$ in LIGO parameter estimation, is constant on the precession timescale $t_{\rm pre}$
\cite{2001PhRvD..64l4013D,2008PhRvD..78d4021R} and is also constant throughout the inspiral to the PN order we consider.
The magnitudes $L$ and $J$ of the orbital and total angular momenta evolve on the radiation-reaction timescale $t_{\rm RR}$.
It is sometimes convenient to define an additional quantity
\begin{equation} \label{E:kappadef}
\kappa \equiv \frac{J^2 - L^2}{2L} = \mathbf{S} \cdot \mathbf{\hat{L}} + \frac{S^2}{2L}~,
\end{equation}
because the limit
\begin{equation} \label{E:kappalim}
\lim_{r \to \infty} \kappa \equiv \kappa_\infty = S_1\cos\theta_{1\infty} + S_2\cos\theta_{2\infty}
\end{equation}
is a finite constant that can be used to label BBHs throughout their inspiral.  In this expression, $\theta_{i\infty}$ is the angle
between $\mathbf{S}_i$ and $\mathbf{L}$ in the limit $r \to \infty$; this angle is a constant since in this limit spin-orbit coupling
dominates over spin-spin coupling and the two spins $\mathbf{S}_i$ simply precess about the orbital angular momentum
$\mathbf{L}$.  The total spin $\mathbf{S}$, as well as the directions of $\mathbf{S}_i$, $\mathbf{L}$, and $\mathbf{J}$ all evolve
on the precession timescale $t_{\rm pre}$.

This last point is somewhat subtle, since in the absence of gravitational radiation, the magnitude and direction of the total
angular momentum $\mathbf{J}$ are both conserved.  In the case of simple precession, $\mathbf{L}$ and $\mathbf{J}$
precess on cones with opening angles $\theta_{Lz}$ and $\theta_J$ respectively about a fixed direction $\mathbf{\hat{z}}$
in an inertial frame \cite{1994PhRvD..49.6274A}.  The timescale hierarchy $t_{\rm pre}/t_{\rm RR} = (r/r_g)^{-3/2} \ll 1$ implies
that $\theta_J/\theta_{Lz} \propto (r/r_g)^{-3/2} \ll 1$, but the frequency $\Omega$ with which $\mathbf{L}$ and $\mathbf{J}$
precess about their cones is the same and of order the inverse of the precession timescale $t_{\rm pre}$.  Although generic
spin precession is more complicated, the direction of the total angular momentum $\mathbf{J}$ still evolves (by a small angle)
on the precession timescale.

\begin{figure}
\includegraphics[width=0.8\columnwidth]{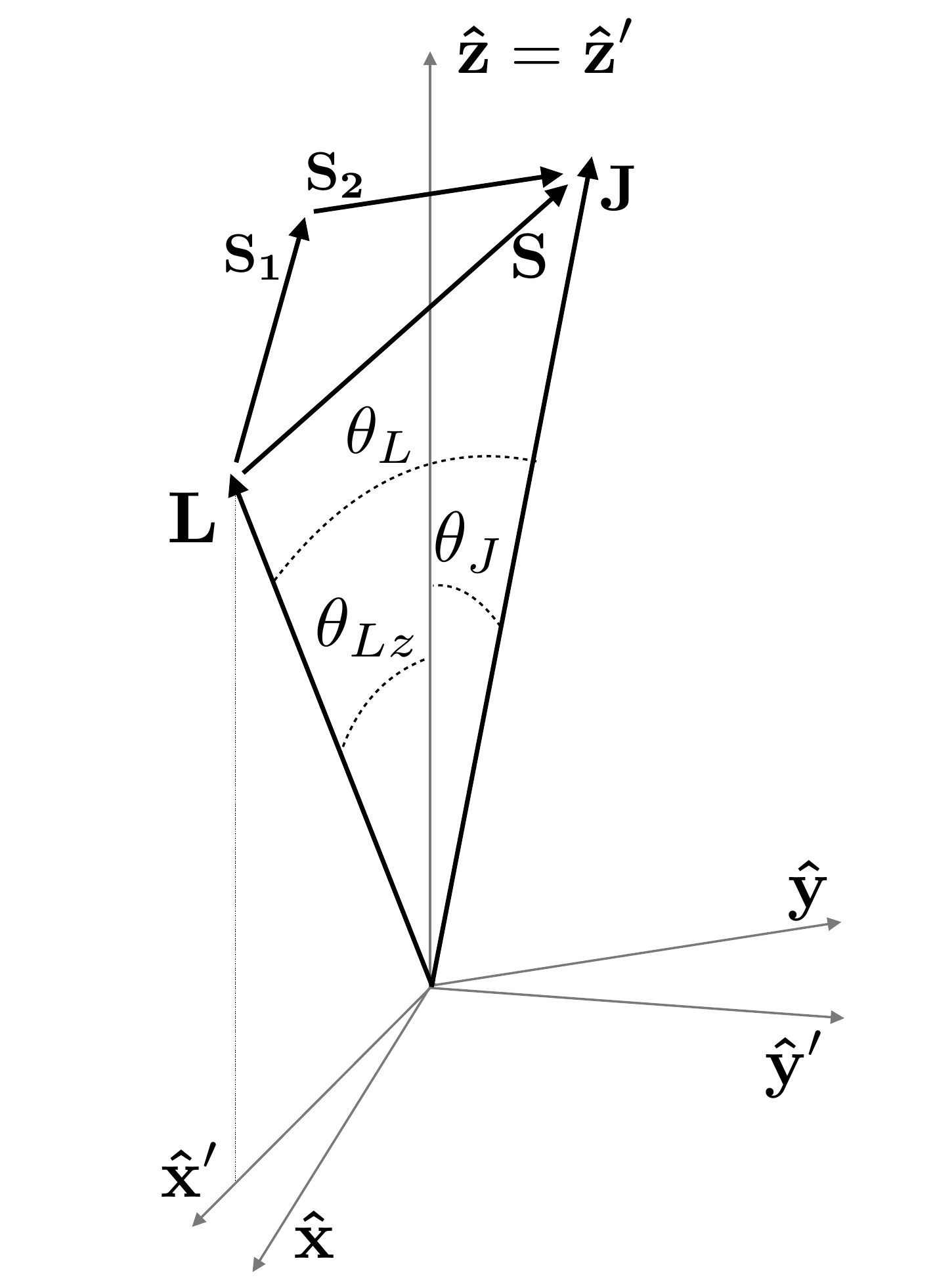}
\caption{References frames useful for describing BBH spin precession.  The total spin $\mathbf{S}$ is the sum of the spins
$\mathbf{S}_1$ and  $\mathbf{S}_2$ of the more massive and less massive black holes.  The total angular momentum
$\mathbf{J}$ is the sum of the orbital angular momentum $\mathbf{L}$ and total spin $\mathbf{S}$, and $\theta_L$ is the angle
between $\mathbf{L}$ and $\mathbf{J}$.  We define the $xyz$ inertial frame such that $\mathbf{\hat{z}}$ points in the direction of
the precession-averaged orbital and total angular momenta $\langle \mathbf{L} \rangle$ and $\langle \mathbf{J} \rangle$ after
many precession cycles.  The basis vectors $\mathbf{\hat{x}}$ and $\mathbf{\hat{y}}$
complete the orthonormal triad.  The angles between $\mathbf{\hat{z}}$ and $\mathbf{L}$ and $\mathbf{J}$ are given by
$\theta_{Lz}$ and $\theta_J$ respectively; the hierarchy $\theta_J \ll \theta_{Lz}$ implies that $\theta_{Lz} \simeq \theta_L$.
After a nutation period $\tau$, $\mathbf{L}$ and $\mathbf{J}$ precess about $\mathbf{\hat{z}}$ by an angle $\alpha$.  We use
these quantities to define the $x'y'z'$ rotating frame in which $\mathbf{\hat{z}'} = \mathbf{\hat{z}}$ and $\mathbf{\hat{x}'}$ and
$\mathbf{\hat{y}'}$ precess about $\mathbf{\hat{z}}$ with precession frequency $\Omega = \alpha/\tau$.} \label{F:frame}
\end{figure}

In previous work \cite{2015PhRvL.114h1103K,2015PhRvD..92f4016G}, we analyzed generic spin precession
under the approximation, valid in the absence of radiation reaction, that the direction of $\mathbf{J}$ stays fixed.  In this
section, we summarize key results from that work which we will use in the following section where we relax the assumption
that the direction of $\mathbf{J}$ stays fixed.  The many constants of motion on the precession timescale listed above imply
that there is only a single degree of freedom in the relative orientations of $\mathbf{L}$, $\mathbf{S}_1$, and $\mathbf{S}_2$,
which we can conveniently specify by choosing the magnitude $S$ of the total spin as a general coordinate.  For precisely equal
masses ($q = 1$), $S$ is constant and an alternative coordinate is required to specify this degree of freedom
\cite{2017CQGra..34f4004G}.  The angle $\theta_L$ between $\mathbf{L}$ and $\mathbf{J}$ is given in terms of $S$ by the
expression
\begin{equation} \label{E:costhetaL}
\cos\theta_L = \frac{J^2 + L^2 - S^2}{2JL}~.
\end{equation}
The hierarchy $\theta_J \ll \theta_{Lz}$ in the PN regime implies that $\theta_{Lz} \simeq \theta_L$ to high accuracy.
The relative orientation of $\mathbf{S}$, $\mathbf{L}$, and $\mathbf{J}$ in terms of this angle are shown in Fig.~\ref{F:frame}.
The total spin magnitude $S$ oscillates in the range $S_- \leq S \leq S_+$, where the extrema $S_\pm$ are the roots of the
equation $\xi = \xi_\pm(S)$, $\xi$ is the projected effective spin given by Eq.~(\ref{E:ProjEff}), and the two curves
\begin{align} \label{E:effpot}
\xi_{\pm}(S) &= \{ (J^2 - L^2 - S^2)[S^2(1+q)^2 - (S_1^2 - S_2^2)(1- q^2)] \notag \\
& \quad \pm (1- q^2) A_1A_2A_3A_4 \}/(4qM^2S^2L)
\end{align}
form a closed loop we called the effective potential for spin precession.  In this expression, we have used four auxiliary functions
$A_i$ which are defined as
\begin{subequations}
\begin{align}
A_1 &\equiv [J^2 - (L - S)^2]^{1/2}\,, \label{E:A1} \\
A_2 &\equiv [(L + S)^2 - J^2]^{1/2}\,, \label{E:A2} \\
A_3 &\equiv [S^2 - (S_1 - S_2)^2]^{1/2}\,, \label{E:A3} \\
A_4 &\equiv [(S_1 + S_2)^2 - S^2]^{1/2}\,. \label{E:A4}
\end{align}
\end{subequations}

Fig.~\ref{F:frame} shows that the oscillations of $S$ correspond to nutation of the orbital angular momentum $\mathbf{L}$,
allowing us to define the nutation period
\begin{equation} \label{E:nutP}
\tau = 2 \int_{S_-}^{S_+}\frac{dS}{|dS/dt|}~.
\end{equation}
and nutation frequency $\omega \equiv 2\pi/\tau$.
Note that in our earlier work \cite{2015PhRvL.114h1103K,2015PhRvD..92f4016G,2016PhRvD..93l4066G}, we referred to $\tau$
as the precession period because we were focused on the relative orientations of the BBH spins and it has the precession
timescale.  The nutation frequency only depends on quantities varying on the radiation-reaction timescale.  The time derivative of
the total spin magnitude $S$ is
\begin{align} \label{E:dSdt}
\frac{dS}{dt} =& -\frac{3(1-q^2)}{2q} \frac{S_1S_2}{S} \frac{(\eta^2M^3)^3}{L^5} \left( 1 - \frac{\eta M^2 \xi}{L} \right) \notag \\
& \times \sin\theta_1 \sin\theta_2 \sin\Delta\Phi~,
\end{align}
where the angles $\theta_i$ between $\mathbf{L}$ and $\mathbf{S}_i$ are given by
\begin{subequations} \label{E:csi}
\begin{align}
\cos\theta_1 &=  \frac{1}{2(1-q)S_1} \left[ \frac{J^2 - L^2 -S^2}{L} - \frac{2qM^2\xi}{1+q} \right]\,, \label{E:cs1} \\
\cos\theta_2 &=  \frac{q}{2(1-q)S_2} \left[ -\frac{J^2 - L^2 -S^2}{L} + \frac{2M^2\xi}{1+q} \right]\,. \label{E:cs2}
\end{align}
\end{subequations}
The angle $\Delta\Phi$ between the projections of $\mathbf{S}_1$ and $\mathbf{S}_2$ orthogonal to $\mathbf{L}$ is given by
\begin{equation} \label{E:DP}
\cos\Delta\Phi = \frac{\cos\theta_{12} - \cos\theta_1\cos\theta_2}{\sin\theta_1\sin\theta_2} \,,
\end{equation}
where
\begin{equation} \label{E:t12}
\cos\theta_{12} = \frac{S^2 - S_1^2 - S_2^2}{2S_1S_2}
\end{equation}
is the cosine of the angle between $\mathbf{S}_1$ and $\mathbf{S}_2$.

Although $\mathbf{L}$, $\mathbf{S}_1$, and $\mathbf{S}_2$ return to their initial relative orientation after a nutation period
$\tau$, in an inertial frame these vectors precess about $\mathbf{\hat{z}}$ by an angle
\begin{equation} \label{E:alpha}
\alpha = 2 \int_{S_-}^{S_+}  \Omega_z \frac{dS}{|dS/dt|}~,
\end{equation}
where
\begin{align} \label{E:Omega_z}
\Omega_z &=\frac{J}{2} \left( \frac{\eta^2M^3}{L^2} \right)^3 \bigg\{ 1 + \frac{3}{2\eta} \left( 1 - \frac{\eta M^2 \xi}{L} \right)
\notag \\ 
&-\frac{3(1+q)}{2qA_1^2A_2^2} \left(1 - \frac{\eta M^2 \xi}{L} \right)[4(1-q)L^2(S_1^2 - S_2^2) \notag \\
&-(1+q)(J^2 - L^2 -S^2)(J^2 - L^2 -S^2 - 4\eta M^2L\xi)] \bigg\}
\end{align}	
is the instantaneous precession frequency.  Note that in our earlier work, we identified $\mathbf{\hat{z}}$ with the instantaneous
direction of the total angular momentum $\mathbf{J}$ rather than its precession average $\langle \mathbf{J} \rangle$, because we
were neglecting the small changes to the direction of $\mathbf{J}$ compared to that of $\mathbf{L}$ ($\theta_J \ll \theta_L$).
These results allow us to define the average precession frequency $\Omega \equiv \alpha/\tau$.  Although the nutation frequency
$\omega$ and precession frequency $\Omega$ are both of order the inverse precession timescale $t_{\rm pre}$, they generally
differ because $\alpha \neq 2\pi$.  As shown in Fig.~\ref{F:frame}, we can define an orthonormal basis for our inertial frame by
choosing vectors $\mathbf{\hat{x}}$ and $\mathbf{\hat{y}}$ perpendicular to $\mathbf{\hat{z}}$.  We can also define a frame
rotating about $\mathbf{\hat{z}} = \mathbf{\hat{z}}'$ with precession frequency $\Omega$ with rotating basis vectors
\begin{subequations} \label{E:rotBV}
\begin{align}
\mathbf{\hat{x}}' &= \mathbf{\hat{x}}\cos\Omega t + \mathbf{\hat{y}}\sin\Omega t~, \\
\mathbf{\hat{y}}' &= -\mathbf{\hat{x}}\sin\Omega t + \mathbf{\hat{y}}\cos\Omega t~.
\end{align}
\end{subequations}	

In the quadrupole approximation, GW emission removes angular momentum from the binary at a rate \cite{1963PhRv..131..435P}
\begin{equation} \label{E:dvecJdt}
\frac{d\mathbf{J}}{dt} = - \frac{32}{5} \left( \frac{r}{M} \right)^{-4} \frac{\eta \mathbf{L}}{M}~.
\end{equation}
The 1PN correction to this expression is also parallel to the orbital angular momentum $\mathbf{L}$
\cite{1995PhRvD..52..821K}.  This expression implies that the magnitudes of $\mathbf{L}$ and $\mathbf{J}$ evolve
according to the equations
\begin{subequations} \label{E:dLdJdt}
\begin{align}
\frac{dL}{dt} &= \frac{d\mathbf{J}}{dt} \cdot \mathbf{\hat{L}} = -\frac{32}{5} \left( \frac{r}{M} \right)^{-4} \frac{\eta L}{M}~,
\label{E:dLdt} \\
\frac{dJ}{dt} &= \frac{d\mathbf{J}}{dt} \cdot \mathbf{\hat{J}} = \frac{dL}{dt} \cos\theta_L~. \label{E:dJdt}
\end{align}
\end{subequations}	
This expression for $dL/dt$ evolves on the radiation-reaction timescale, but the expression for $dJ/dt$ evolves on the
precession timescale because of the angular term $\cos\theta_L$ given by Eq.~(\ref{E:costhetaL}).  We can precession
average the right-hand side of Eq.~(\ref{E:dJdt}) using
\begin{equation}
\langle \cos\theta_L \rangle = \frac{2}{\tau} \int_{S_-}^{S_+}\frac{\cos\theta_L~dS}{|dS/dt|}
\end{equation}
to obtain the precession-averaged loss of total angular momentum
$\langle dJ/dt \rangle = (dL/dt) \langle \cos\theta_L \rangle$ \cite{2015PhRvL.114h1103K,2015PhRvD..92f4016G}.  This
equation and Eq.~(\ref{E:dLdt}) can be numerically integrated with a time step on the radiation-reaction timescale,
providing a vast savings in computational time compared to a time step on the precession timescale if one is only interested
in the relative orientations of $\mathbf{L}$, $\mathbf{S}_1$, and $\mathbf{S}_2$ specified by Eqs.~(\ref{E:csi}) and (\ref{E:DP})
\cite{2015PhRvD..92f4016G,2016PhRvD..93l4066G}.  However, to determine the directions of the vectors $\mathbf{L}$ and
$\mathbf{J}$ in an inertial frame (perhaps for the purpose of calculating the emission of GWs), one must integrate the
instantaneous precession frequency $\Omega_z$ given by Eq.~(\ref{E:Omega_z}) with a time step on the precession timescale.
In the next section, we derive new series expansions for $\mathbf{L}$ and $\mathbf{J}$ in terms of quantities that only evolve on
the radiation-reaction timescale which can in principle achieve similar computational savings to our earlier expression for
$\langle dJ/dt \rangle$.

\section{A New Expansion} \label{S:expansion}

In the inertial (unprimed) frame defined in the previous section, we can decompose the orbital angular momentum
\begin{equation} \label{E:PAorb}
\mathbf{L} = L_\parallel \mathbf{\hat{z}} + \mathbf{L}_\perp
\end{equation}
into components parallel and perpendicular to the direction $\mathbf{\hat{z}}$ of its precession average
$\langle \mathbf{L} \rangle$.  Without loss of generality, we can choose $\mathbf{L}$ to lie in the $xz$ plane at $t=0$ with total
spin magnitude $S = S_-$.  With this choice, the perpendicular component of $\mathbf{L}$ is given by
\begin{equation} \label{E:Lin}
\mathbf{L}_\perp = L\sin\theta_L(\cos\Phi_L\mathbf{\hat{x}} + \sin\Phi_L\mathbf{\hat{y}})~.
\end{equation}	
The total spin magnitude $S(t)$ is an even function of time with period $\tau$, implying that it is fully specified by its values
in the interval $0 \leq t \leq \tau/2$.  On this interval, $S(t)$ is the inverse of the function
\begin{equation} \label{E:tofS}
t(S) = \int_{S_-}^{S} \frac{dS'}{|dS/dt|}~,
\end{equation}
where $dS/dt$ is given by Eq.~(\ref{E:dSdt}) and $S_-\leq S \leq S_+$.  Eq.~(\ref{E:costhetaL}) indicates that $\theta_L$ is similarly a periodic,
even function of time, while Eq.~(\ref{E:Omega_z}) requires $\Phi_L(t)$ to be a periodic, odd function of time defined by its
values
\begin{equation} \label{E:PhiL_TD}
\Phi_L(t) = \int_0^t \Omega_z~dt' = \int_{S_-}^{S}  \Omega_z \frac{dS'}{|dS/dt|}
\end{equation}
in the interval $0 \leq t \leq \tau/2$.  The symmetry and periodicity of $\theta_L$ and $\Phi_L$ imply that we can Fourier
expand the perpendicular component of $\mathbf{L}$ in the rotating frame given by Eq.~(\ref{E:rotBV}) to obtain the series
\begin{equation} \label{E:rotbas}
\mathbf{L}_\perp(t) = L\sum_{n=0}^{+\infty} [\theta_{Lxn}' \cos (n\omega t)~\mathbf{\hat{x}}'
+ \theta_{Lyn}' \sin (n\omega t)~\mathbf{\hat{y}}' ]~.
\end{equation}	
Comparing Eqs.~(\ref{E:Lin}) and (\ref{E:rotbas}) and using Eq.~(\ref{E:rotBV}) to relate the rotating and inertial frames, we see that 
\begin{align}
\theta_{Lxn}' &= \frac{2 - \delta_{n0}}{L\tau} \int_0^\tau \mathbf{L}_\perp \cdot \mathbf{\hat{x}}'\cos (n\omega t)~dt \notag \\
&= \frac{2}{\tau} \int_{S_-}^{S_+} \cos(\Phi_L - \Omega t) \sin\theta_L\cos (n\omega t) \frac{dS}{|dS/dt|} \label{E:tLxn'}
\end{align}
and
\begin{align}
\theta_{Lyn}' &= \frac{2 - \delta_{n0}}{L\tau} \int_0^\tau \mathbf{L}_\perp \cdot \mathbf{\hat{y}}'\sin (n\omega t)~dt \notag \\
&= \frac{4}{\tau} \int_{S_-}^{S_+} \sin(\Phi_L - \Omega t) \sin\theta_L\sin (n\omega t) \frac{dS}{|dS/dt|}~, \label{E:tLyn'}
\end{align}
where the Kronecker delta $\delta_{ij}$ equals unity for $i=j$ and zero otherwise.
We can use Eqs.~(\ref{E:rotBV}) and (\ref{E:rotbas}) to obtain an equivalent series for $\mathbf{L}_\perp$ in the inertial frame,
\begin{equation} \label{E:inbas}
\mathbf{L}_\perp(t) = L\sum_{n=-\infty}^{+\infty} \theta_{Ln} \{ \cos[(\Omega - n\omega)t] \mathbf{\hat{x}} +
\sin[(\Omega - n\omega)t] \mathbf{\hat{y}} \}~,
\end{equation}
where
\begin{align}
\theta_{Ln} &= \frac{1 + \delta_{n0}}{2}[\theta_{Lx|n|}' -  {\rm sgn}(n)\theta_{Ly|n|}']  \notag \\
&= \frac{2}{\tau} \int_{S_-}^{S_+} \cos(\Phi_L - \Omega t + n\omega t) \sin\theta_L\frac{dS}{|dS/dt|}~. \label{E:tLn} 
\end{align}
One can obtain $\mathbf{L}$ from Eqs.~(\ref{E:PAorb}) and (\ref{E:inbas}) by recognizing that the magnitude of $\mathbf{L}$ is
conserved on the precession timescale implying that 
\begin{equation}
L_\parallel = \sqrt{L^2 - \mathbf{L}_\perp \cdot \mathbf{L}_\perp}~.
\end{equation}

\begin{figure*}
\includegraphics[width=2\columnwidth]{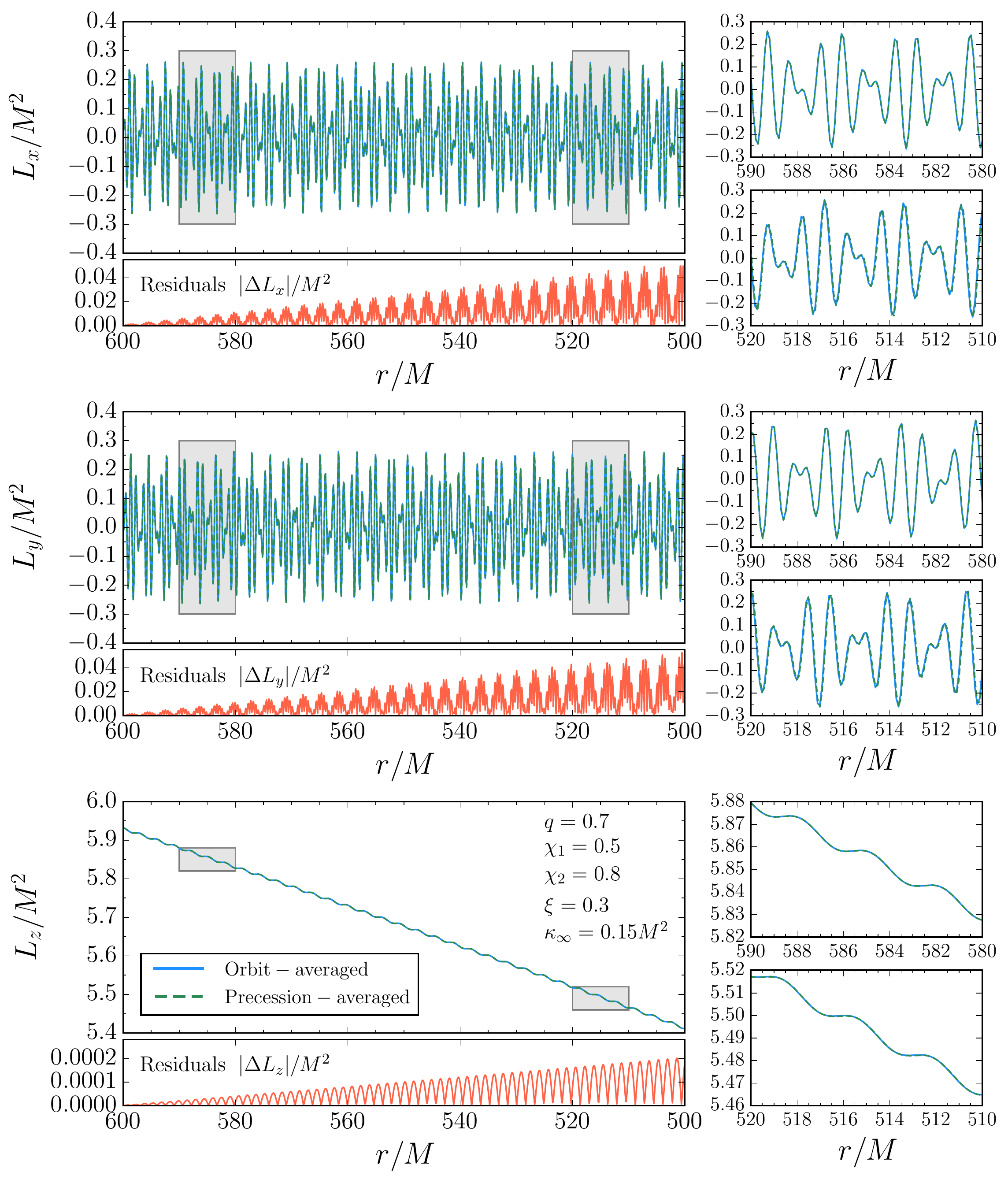}
\caption{Comparison of the orbital angular momentum $L(t)$ determined from a numerical integration of the orbit-averaged
spin-precession equations and our new precession-averaged series expansion given by Eq.~(\ref{E:inbas}).  The binary has
mass ratio $q = 0.7$, dimensionless spin magnitudes $\chi_1 = 0.5$ and $\chi_2 = 0.8$, projected effective spin $\xi = 0.3$, and
asymptotic projected total spin $\kappa_\infty = 0.15 M^2$.  It inspirals from binary separation $r = 600 M$ to $r = 500 M$.  The top,
middle, and bottom panels show the components of $\mathbf{L}$ in an inertial frame in which $\mathbf{\hat{z}}$ points in the
direction of $\langle \mathbf{L} \rangle$, the precession-averaged orbital angular momentum.  The solid
blue curves show the orbit-averaged solution, the dashed green curves show our new precession-averaged solution, and the
red curves below each panel show the magnitude of the differences between the solutions.  We have only used the five terms
corresponding to $n=0, \pm 1, \pm 2$ in the series expansion.  The right panels show zoomed-in views of the inspiral from
$r = 590 M$ to $r = 580 M$ and $r = 520 M$ to $r = 510 M$.} \label{F:comp}
\end{figure*}

Eq.~(\ref{E:inbas}) is an elegant expression for $\mathbf{L}_\perp$; each term corresponds to a vector with magnitude
$L|\theta_{Ln}|$ tracing out a circle with frequency $\Omega - n\omega$ in the plane orthogonal to $\mathbf{\hat{z}}$, the direction of
the precession-averaged orbital angular momentum $\langle \mathbf{L} \rangle$.  These magnitudes and frequencies are both
evolving on the radiation-reaction timescale $t_{\rm RR}$, implying that they can be numerically evaluated throughout the inspiral
with a time step of order $t_{\rm RR}$ leading to potentially large computational savings.  Eq.~(\ref{E:inbas}) also seems well suited
for Fourier transformation if one is interested in functions in the frequency domain for GW analysis.  We test its validity by comparing
it to numerical integration of the full spin-precession equations.  We show this comparison in Fig.~\ref{F:comp}, including only the
$n = 0$, $\pm 1$, and $\pm 2$ terms in Eq.~(\ref{E:inbas}).  As we are allowing the binary to inspiral while making the comparison,
we must replace the arguments $(\Omega - n\omega)t$ of the sinusoids in Eq.~(\ref{E:inbas}) by the phases
\begin{equation} \label{E:phaseRR}
\psi_n(t) = \int_0^t (\Omega - n\omega)~dt'~.
\end{equation}
We see excellent agreement between our new precession-averaged series expansion in Eq.~(\ref{E:inbas}) and the
traditional numerical solutions of the orbit-averaged precession equations, shown respectively by the green dashed and solid
blue curves.
The $z$-component of $\mathbf{L}$ (in the direction of its precession-averaged value) calculated in the two approaches agrees to
a part in $10^4$, while residuals for the perpendicular component $\mathbf{L}_\perp$ grow to about the 1\% level by the time the
binary inspirals from $600 M$ to $500 M$.
These residuals result from numerical error in the phasing given by Eq.~(\ref{E:phaseRR}); the neglected terms with $|n| \geq 3$
remain highly subdominant.

Although the dashed green curves in Fig.~\ref{F:comp} include five terms from Eq.~(\ref{E:inbas}), the precessional modulation
seen in this figure results from just two dominant terms of nearly equal magnitude.  For most of the inspiral, these two terms are
the $n = -1$ and $n = 0$ terms in the expansion of Eq.~(\ref{E:inbas}), but for a discrete interval between $r = 600 M$ and
$r = 500 M$, the two dominant terms are instead $n = 0$ and $n = +1$.  This results not from the continuous evolution of the
coefficients $\theta_{Ln}$ on the radiation-reaction timescale, but from two discontinuities.  At two points during the inspiral from
$r = 600 M$ to $r = 500 M$, the magnitudes of the orbital and total angular momentum $L$ and $J$ attain values such that
$\mathbf{L}$ and $\mathbf{J}$ are instantaneously aligned once per nutation period at $S = S_-$.  This alignment implies that
$\alpha$, the angle by which $\mathbf{L}$ precesses about $\mathbf{J}$ over a nutation period, cannot be defined
\cite{2016PhRvD..93d4071T}.  This is purely a coordinate issue, analogous to the inability to define the total change in longitude
on a trip that passes directly over the North Pole.  When an inspiraling binary passes through values of $L$, $J$, and $\xi$ for
which alignment between $\mathbf{L}$ and $\mathbf{J}$ is possible, $\alpha$ changes discontinuously by $\pm2\pi$ implying that
the precession frequency $\Omega = \alpha/\tau$ changes discontinuously by the nutation frequency $\omega = 2\pi/\tau$.  A shift
$\Omega \to \Omega' = \Omega \pm \omega$ leads to a shift $\theta_{Ln} \to \theta'_{Ln} = \theta_{L,n\mp1}$ according to
Eq.~(\ref{E:tLn}).  This shift will leave the infinite summation in Eq.~(\ref{E:inbas}) unchanged, merely relabeling the individual
terms.  Such shifts occur twice during the inspiral from $r = 600 M$ to $r = 500 M$ of the binary shown in Fig.~\ref{F:comp};
$\alpha$ first increases by $2\pi$, shifting the dominant terms from $n = \{ -1, 0 \}$ to $n =  \{ 0, +1 \}$, then decreases by $2\pi$,
restoring $n = \{-1, 0\}$ as the dominant terms.  The summation of the five terms $n = \{ 0, \pm1, \pm2 \}$ shown in Fig.~\ref{F:comp}
always includes the two dominant terms and thus leaves no observable discontinuities in $\mathbf{L}$.

\begin{figure}
\includegraphics[width=1.0\columnwidth]{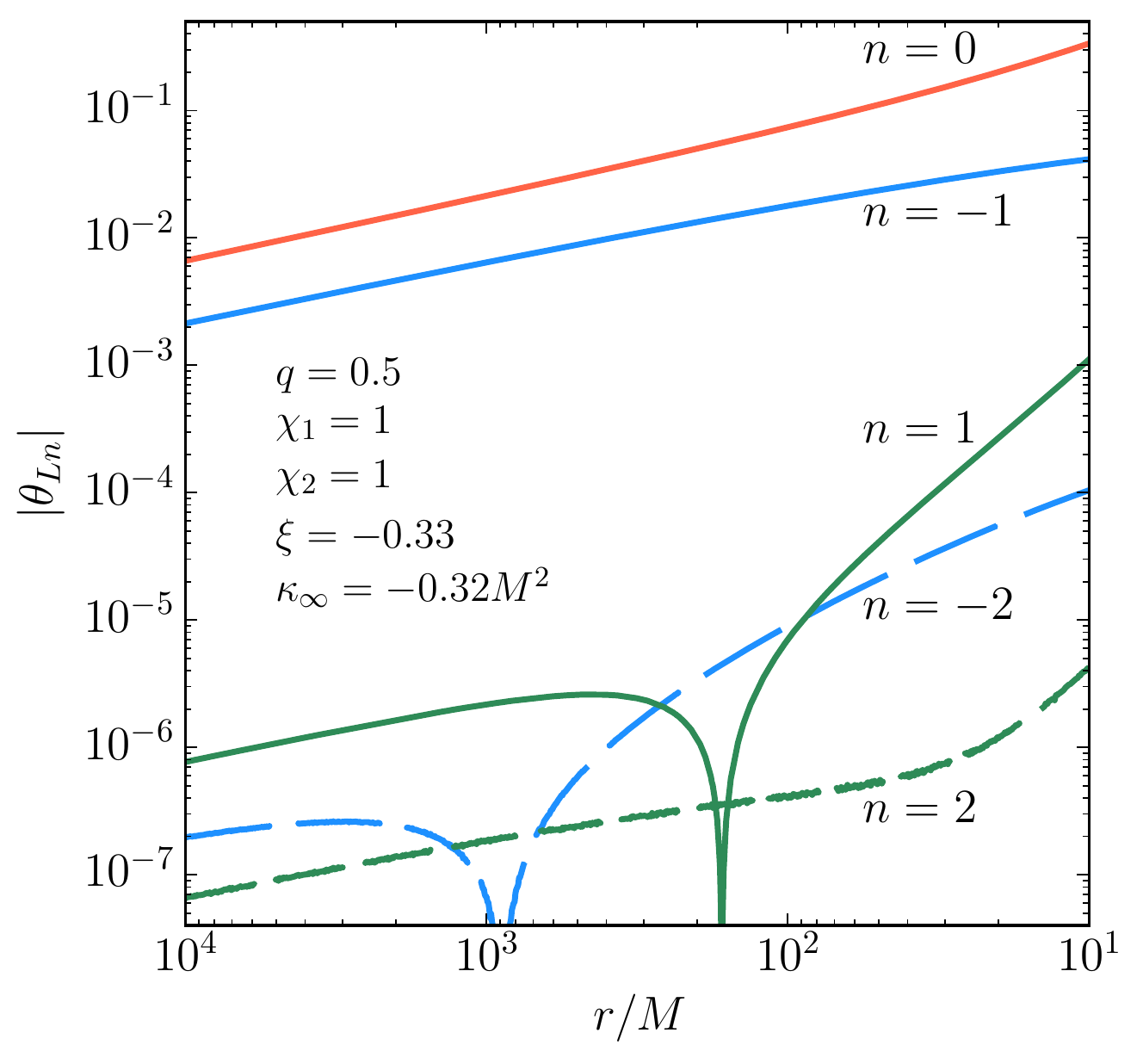}
\caption{The magnitudes of the five largest coefficients $\theta_{Ln}$ in the series expansion of Eq.~(\ref{E:inbas}) as a function
of binary separation $r$ for a binary with mass ratio $q = 0.5$, maximal dimensionless spins $\chi_1 = \chi_2 = 1$, projected
effective spin $\xi = -0.33$, and asymptotic projected total spin $\kappa_\infty = -0.32 M^2$.  The $n=0$ and $n=-1$ coefficients
shown by the solid red and blue curves remain the two largest coefficients throughout the inspiral, while the $n=+1,+2,-2$
coefficients, shown by the solid green, dashed green, and dashed blue curves respectively, remain less than 1\% of the two
dominant curves until just before the final separation $r = 10M$.} \label{F:coeff}
\end{figure}

We show the five largest coefficients $\theta_{Ln}$ for $n = 0, \pm 1, \pm 2$ during the inspiral of a different binary from
$r = 10^4 M$ to $10 M$ in Fig.~\ref{F:coeff}.  The parameters for this binary, listed in the caption to the figure, were chosen such
that the binary passes through a nutational resonance at $r \simeq 481 M$.  Such nutational resonances are the focus of
Sections~\ref{S:resonances} and \ref{S:demo}; the same binary is also shown in Figs.~\ref{F:Jperp_terms} and \ref{F:Euler}.  This
binary differs from the one shown in Fig.~\ref{F:comp} in that it does not pass through any discontinuities in $\alpha$, but shares
the common feature that the $n = -1, 0$ terms are dominant through most of the inspiral.  For such binaries, the precession of
$\mathbf{L}$ can be modeled to $\sim1\%$ accuracy using just the two dominant terms in Eq.~(\ref{E:inbas}) whose coefficients
vary smoothly on the radiation-reaction timescale.  This suggests that precession averaging can provide computational savings
for the evolution of $\mathbf{L}$ during an inspiral similar to those obtained for the evolution of the total angular momentum $J$
demonstrated in our previous work \cite{2015PhRvL.114h1103K,2015PhRvD..92f4016G}.

Our new expansion in Eq.~(\ref{E:inbas}) can also be used to calculate the evolution of the total angular momentum
$\mathbf{J}(t)$ in our inertial $xyz$ frame.  If the rate at which angular momentum is radiated is related to the orbital angular
momentum by Eq.~(\ref{E:dvecJdt}), our expansion implies that 
\begin{align} \label{E:dJperpdt}
\frac{d\mathbf{J}_\perp}{dt} &= -\frac{32}{5} \left( \frac{r}{M} \right)^{-4} \frac{\eta\mathbf{L}_\perp}{M} \notag \\
&= \frac{dL}{dt} \sum_{n=-\infty}^{+\infty} \theta_{Ln}  \{ \cos[(\Omega - n\omega)t] \mathbf{\hat{x}}
+ \sin[(\Omega - n\omega)t] \mathbf{\hat{y}} \}~.
\end{align}
If we integrate this expression on the precession timescale, holding fixed the amplitudes and frequencies varying on the longer
radiation-reaction timescale, we find a similar expansion for the perpendicular component of the total angular momentum,
\begin{equation} \label{E:JperpExp}
\mathbf{J}_\perp(t) = J\sum_{n=-\infty}^{+\infty} \theta_{Jn} \{ \sin[(\Omega - n\omega)t] \mathbf{\hat{x}} -
\cos[(\Omega - n\omega)t] \mathbf{\hat{y}} \}~,
\end{equation}
where the coefficients in the two expansions of Eqs.~(\ref{E:inbas}) and (\ref{E:JperpExp}) are proportional to each other:
\begin{equation} \label{E:thetaJn}
\frac{\theta_{Jn}}{\theta_{Ln}} = \frac{1}{J} \frac{dL}{dt} \left( \frac{1}{\Omega - n\omega} \right) \propto
\frac{t_{\rm pre}}{t_{\rm RR}} \propto \left( \frac{r}{M} \right)^{-3/2}.
\end{equation}
This agrees with the earlier finding that for simple precession, the total angular momentum $\mathbf{J}$ precesses about
a cone with opening angle $\theta_J \propto (r/M)^{-2}$ much less than the opening angle $\theta_L \propto (r/M)^{-1/2}$ of the
cone about which the orbital angular momentum $\mathbf{L}$ precesses \cite{1994PhRvD..49.6274A}.
Eq.~(\ref{E:thetaJn}) reveals that $\theta_{Jn}$ diverges for $\Omega = n\omega$, mathematically equivalent to
$\alpha = 2\pi n$ from our definitions of the precession and nutation frequencies in Sec.~\ref{S:review}.  This condition, which
we call a nutational resonance, has potentially profound implications for the evolution of $\mathbf{J}$ which we explore in the
next section.

\section{Nutational Resonances} \label{S:resonances}

At a nutational resonance, the arguments of the sinusoids in the $n = \Omega/\omega$ term in Eq.~(\ref{E:dJperpdt})
vanish, implying that this term corresponds to constant emission of angular momentum in the $x$ direction.  This emission
will cause the precession-averaged total angular momentum $\langle \mathbf{J} \rangle$ to tilt towards the $x$ axis and away
from its initial direction which defined the $z$ axis.  This tilting behavior will not continue indefinitely, because the precession
frequency $\Omega$ and nutation frequency $\omega$ are both evolving on the radiation-reaction timescale $t_{\rm RR}$.  A
generic binary will not be in a nutational resonance ($\Omega/\omega$ will not be an integer), but as it inspirals towards merger it
may pass through one or more of such resonances.  At each passage through a nutational resonance, the precession-averaged
total angular momentum $\langle \mathbf{J} \rangle$ will tilt by some angle $\theta_{\rm tilt}$, providing a randomly oriented ``kick''
of magnitude $J\theta_{\rm tilt}$ to $\mathbf{J}_\perp$ in an inertial frame.  These kicks will accumulate throughout the inspiral
causing $\langle \mathbf{J} \rangle$ to random walk away from its initial direction at large separations set by binary formation. 
Whether these tilts are astrophysically relevant or lead to detectable GW signatures depends on both the magnitudes of the tilt
angles $\theta_{\rm tilt}$ and the frequency with which binaries encounter nutational resonances.  We will derive an analytic
estimate of the tilt angle $\theta_{\rm tilt}$ in this section, then use this estimate to explore the distribution of tilt angles as a function
of binary parameters in Sec.~\ref{S:demo}.
		
\begin{figure*}
\includegraphics[width=2\columnwidth]{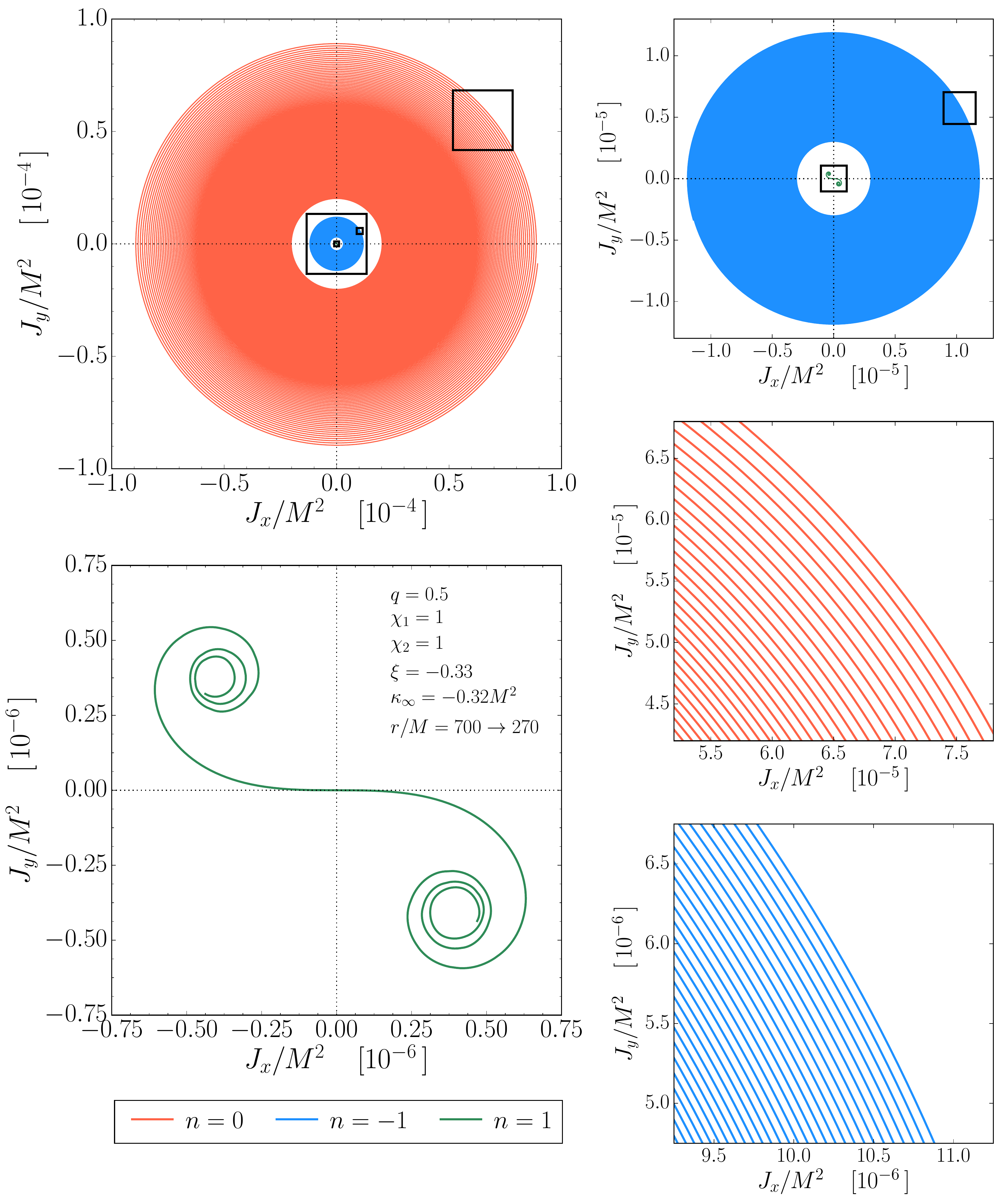}
\caption{The evolution of $\mathbf{J}_\perp$, the component of the total angular momentum in the $xy$ plane, as binary
black holes with mass ratio $q = 0.5$, maximal dimensionless spins $\chi_1 = \chi_2 = 1$, and projected effective spin
$\xi = -0.33$ inspiral from $r = 700 M$ to $r = 270 M$.  The binary encounters a $n = 1$ nutational resonance
($\Omega = \omega$) at $r \simeq 481 M$ when the magnitude of the total angular momentum $J \simeq 4.56 M$.  The direction
of $\mathbf{J}$ at this point in the inspiral defines the $z$ axis.  The red, blue, and green curves show numerical integration of
the $n = \{ 0, -1, +1 \}$ terms respectively in the expansion of Eq.~(\ref{E:dJperpdt}).  The top left panel shows the largest view,
while the top right and middle right panels show two insets to this figure.  The bottom left and bottom right panels show the
two insets to the top right panel.  The bottom left panel most clearly shows that the resonant $n = 1$ term shown by the green
curve is tilted as it passes through resonance.} \label{F:Jperp_terms}
\end{figure*}

We show an example of BBHs passing through a $n = 1$ nutational resonance in Fig.~\ref{F:Jperp_terms}.  We integrate
Eq.~(\ref{E:dJperpdt}) numerically backwards and forwards in time from the resonance at $r \simeq 481 M$, defining the $z$ axis
to point in the direction of the precession-averaged total angular momentum $\langle \mathbf{J} \rangle$ at this binary separation.
We show the dominant non-resonant $n = 0$ and $n = -1$ terms with red and blue curves respectively, while the resonant
$n = 1$ term is shown by the green curve.  On the precession timescale, the non-resonant $n = 0$ and $n = -1$ terms trace circles
in the $xy$ plane with radii $J|\theta_{J0}|$ and $J|\theta_{J,-1}|$ and frequencies $\Omega$ and $\Omega + \omega$, consistent
with the expansion for $\mathbf{J}_\perp$ given in Eq.~(\ref{E:JperpExp}).  On the radiation-reaction timescale, these curves spiral
outwards as $\theta_{Jn}$ increase in magnitude as the binary separation $r$ decreases from $700 M$ to $270 M$.

The resonant $n = 1$ term exhibits qualitatively different behavior, in addition to being much smaller in magnitude consistent with
the hierarchy of coefficients shown in Fig.~\ref{F:coeff}.  At large separations, where the precession frequency $\Omega$ and
nutation frequency $\omega$ have not quite achieved resonance, the $n = 1$ term precesses in small circles with radii
$J|\theta_{J1}|$ and very small frequency $\Omega - \omega$.  This is shown in the top left corner of the bottom left panel of
Fig.~\ref{F:Jperp_terms}.  As the binary approaches resonance, the angular momentum loss due to this term comes to point in a fixed
direction on the precession timescale (along the $x$ axis).  Next, the binary passes through resonance when the green curve
reaches the origin at $J_x = J_y = 0$.  Finally, the $n = 1$ term resumes precession with frequency $\Omega - \omega$ (now
negative) along circles with radii $J|\theta_{J1}|$ as shown in the bottom right corner of the bottom left panel of
Fig.~\ref{F:Jperp_terms}.  The axes about which the $n = 1$ term precesses before and after resonance are displaced with respect to
each other, corresponding to a tilt in the precession-averaged total angular momentum $\langle \mathbf{J} \rangle$.

\begin{figure}
\includegraphics[width=1\columnwidth]{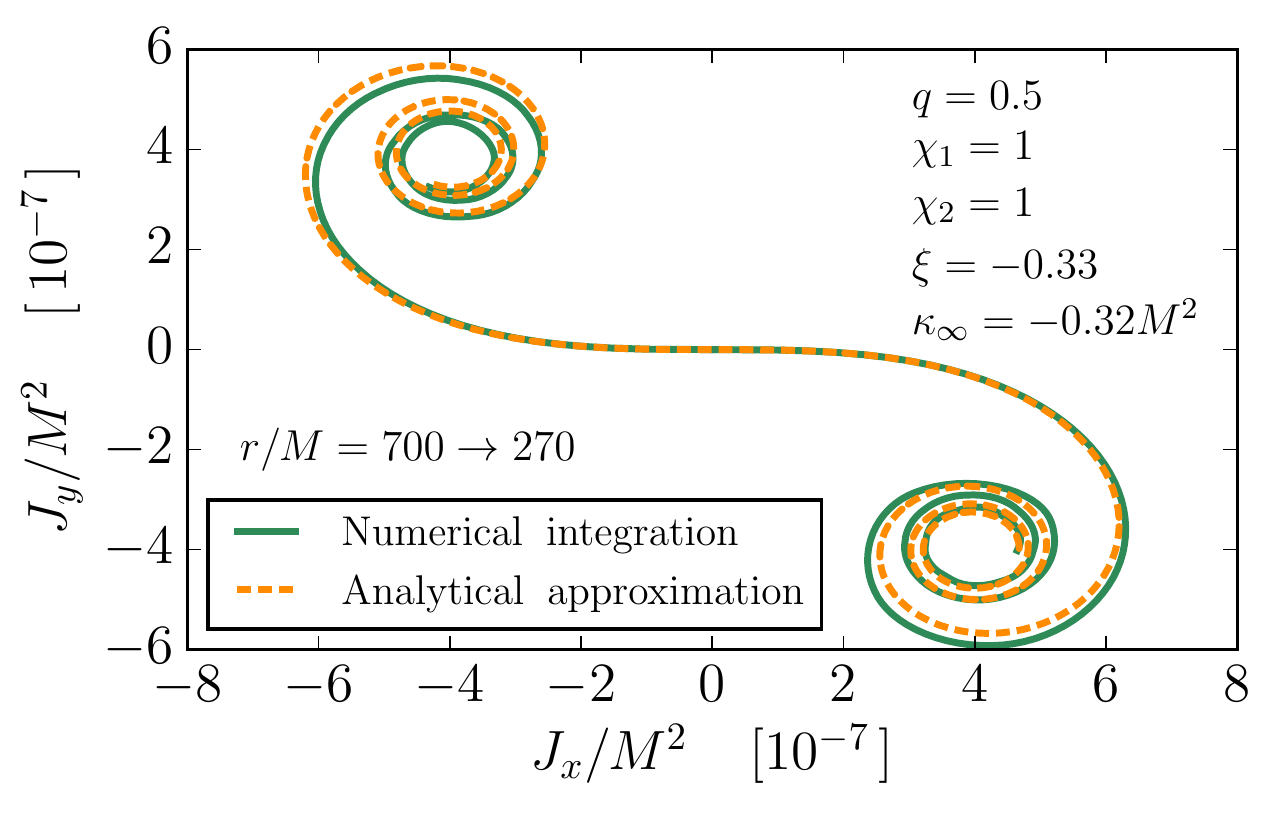}
\caption{A comparison between numerical integration of the resonant term in Eq.~(\ref{E:dJperpdt}) for the nutational resonance
depicted in Fig.~\ref{F:Jperp_terms} and our analytical approximation given by Eq.~(\ref{E:Jperp_res}).  The agreement is
excellent; the symmetric Euler spiral shown by the dashed orange curve nearly perfectly describes the numerical integration
shown by the solid green curve despite the significant changes in $L$ and $J$ as the binary inspirals from $r = 700 M$ to
$r = 270 M$.} \label{F:Euler}
\end{figure}

We can estimate the magnitude of this tilt by Taylor expanding the resonant term in Eq.~(\ref{E:dJperpdt}) about the resonance
and integrating analytically.  We begin with the frequency of the resonant term	,
\begin{equation} \label{E:freqT} 
\Omega - n\omega \simeq \left[ \frac{d(\Omega - n\omega)}{dL} \right]_0 (L - L_0) = sD^2t~,
\end{equation}
where the total derivative of the frequency with respect to the magnitude of the orbital angular momentum is evaluated at
resonance where $L = L_0$.  In this expression, we have also defined the binary to pass through resonance at $t = 0$ and two
constants
\begin{align}
s &\equiv {\rm sgn}\left( \frac{d\alpha}{dL} \frac{dL}{dt} \frac{1}{\tau} \right)~, \label{E:sdef} \\
D &\equiv \left| \frac{d\alpha}{dL} \frac{dL}{dt} \frac{1}{\tau} \right|^{1/2} \propto \frac{1}{\sqrt{t_{\rm RR}t_{\rm pre}}} \propto
\left( \frac{r}{M} \right)^{-13/4}. \label{E:Ddef}	
\end{align}
Eqs.~(\ref{E:freqT}) and (\ref{E:phaseRR}) imply that the phase near resonance is given by
\begin{equation} \label{E:psiT}  
\psi_n = \int_0^t (\Omega - n\omega)~dt' \simeq \frac{1}{2}sD^2t^2~.
\end{equation}
Inserting this phase into the arguments of the sinusoids of the resonant term in Eq~(\ref{E:dJperpdt}), we find that 	
\begin{equation} \label{E:dJperpdt_res}
\frac{d\mathbf{J}_{\perp n}}{dt} = \frac{dL}{dt} \theta_{Ln} \left[  \cos\left( \frac{1}{2}D^2t^2 \right)
\mathbf{\hat{x}} + s \sin\left( \frac{1}{2}D^2t^2 \right) \mathbf{\hat{y}} \right]
\end{equation}
Integrating Eq.~(\ref{E:dJperpdt_res}) leads to	
\begin{equation} \label{E:Jperp_res}
\mathbf{J}_{\perp n} = \frac{\sqrt{2}}{D} \frac{dL}{dt} \theta_{Ln} \left[  \mathcal{C}\left( \frac{Dt} {\sqrt{2}}	\right) \mathbf{\hat{x}}
+ s \mathcal{S}\left( \frac{Dt}{\sqrt{2}}  \right) \mathbf{\hat{y}} \right]
\end{equation}	
where $\mathcal{C}(x)$ and $\mathcal{S}(x)$ are the Fresnel integrals
\begin{subequations} \label{E:FresInt}
\begin{align}
\mathcal{C}(x) & \equiv \int_0^x \cos t^2~dt~, \\
\mathcal{S}(x) & \equiv \int_0^x \sin t^2~dt~.
\end{align}
\end{subequations}
Eq.~(\ref{E:Jperp_res}) indicates that the resonant term $\mathbf{J}_{\perp n}$ can be approximated as an Euler spiral.  We
compare this Euler spiral to a numerical integration of the resonant term in Eq.~(\ref{E:dJperpdt}) in Fig.~\ref{F:Euler}.

The Fresnel integrals have limiting values
\begin{equation}
\lim_{x\to\pm\infty} \mathcal{C}(x), \mathcal{S}(x) = \pm\sqrt{\pi/8}
\end{equation}
which allow us to estimate the total shift
\begin{equation} \label{E:Jshift}
\Delta \mathbf{J}_{\perp n} \equiv \mathbf{J}_{\perp n}(\infty) - \mathbf{J}_{\perp n}(-\infty)
\end{equation}
in the precession-averaged total angular momentum relative to its direction at resonance as a binary passes through a nutational
resonance.  This in turn implies that $\mathbf{J}$ tilts by an angle
\begin{align} \label{E:thetatilt}
\theta_{\rm tilt} &= \frac{|\Delta \mathbf{J}_{\perp n}|}{J} = \frac{(2\pi)^{1/2}}{JD} \frac{dL}{dt}\theta_{Ln}
\notag \\
& \propto \left( \frac{t_{\rm {pre}}}{t_{\rm {RR}}} \right)^{1/2} \theta_{Ln} \propto \left( \frac{r}{M} \right)^{-5/4}~.
\end{align}
For the nutational resonance shown in Fig.~\ref{F:Euler}, the total shift $\Delta \mathbf{J}_{\perp n}$ predicted by 
Eq.~(\ref{E:Jshift}) agrees with the numerical result obtained by integrating Eq.~(\ref{E:dJperpdt}) to better than 1\%.  This justifies
our use of Eq.~(\ref{E:thetatilt}) in the next section to estimate how the precession-averaged total angular momentum
$\langle \mathbf{J} \rangle$ tilts as BBHs encounter nutational resonances during their inspirals.
		
\section{Distribution of Nutational Resonances} \label{S:demo}

\begin{figure*}
\includegraphics[width=2\columnwidth]{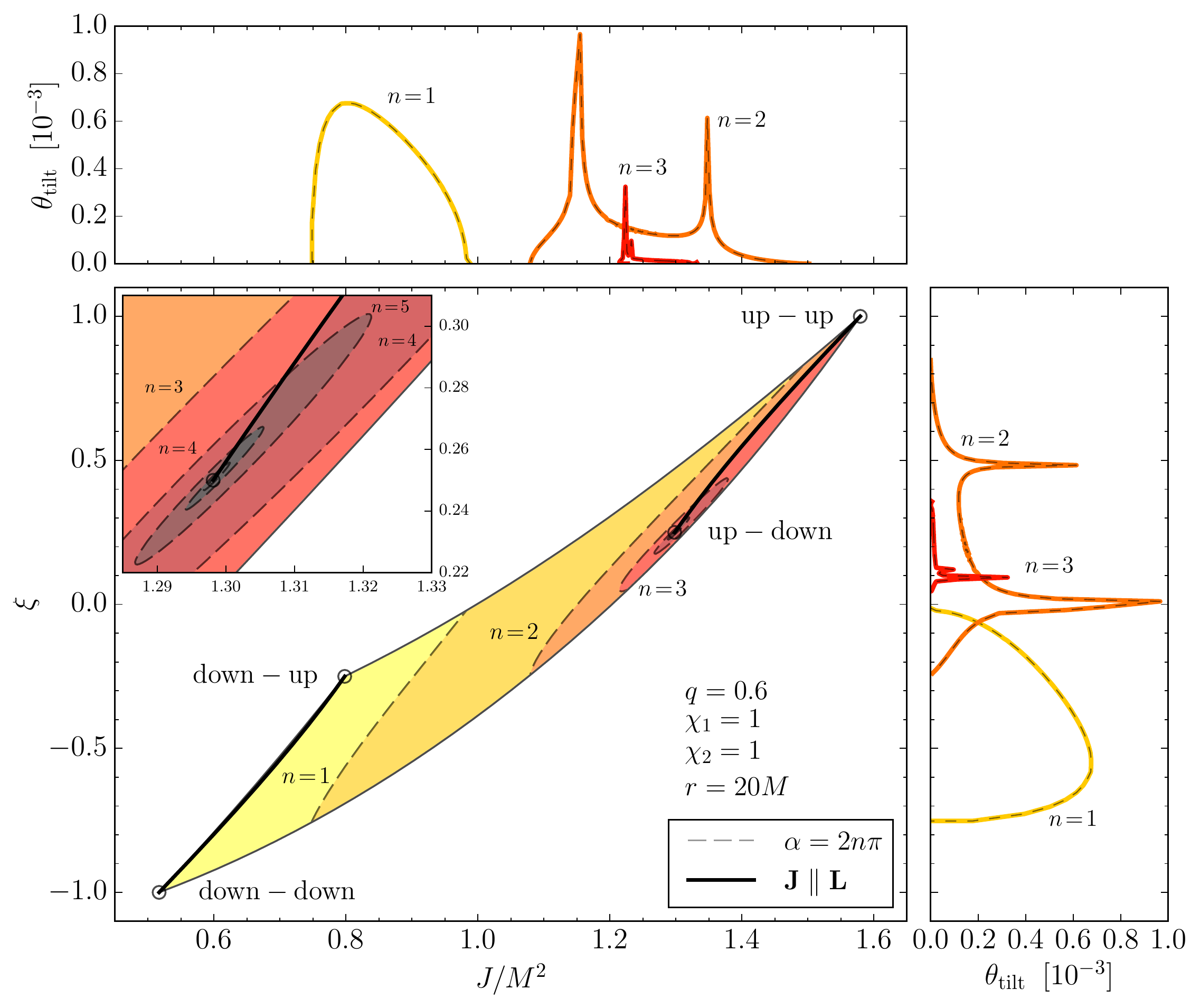}
\caption{{\it Central panel:} A contour plot of the precession angle $\alpha$ as a function of the magnitude of the total angular
momentum $J$ and the projected effective spin $\xi$ for BBHs with mass ratio $q = 0.6$, dimensionless spin magnitudes
$\chi_1 = \chi_2 = 1$, and binary separation $r = 20 M$.  The colored region shows the allowed values of $J$ and $\xi$ for these
parameters, with the three cusps on the boundaries corresponding to the up-up, down-up, and down-down configurations.  The
thin dashed lines indicate nutational resonances ($\alpha = 2\pi n$ for integer $n$).  The inset shows parameter space near the
unstable up-down configuration where $\alpha \to \infty$.  The solid black lines show binaries for which $\mathbf{J}$ and the
orbital angular momentum $\mathbf{L}$ are parallel at $S = S_-$, implying that $\alpha$ is undefined.  The value of $\alpha$
along contours crossing these lines changes by $\pm 2\pi$.  {\it Top and right panels:} The tilt angle $\theta_{\rm tilt}$ along the
$\alpha = 2\pi n$ nutational resonance contours as functions of $J$ and $\xi$ respectively.  The spikes on these curves
correspond to parameters for which $d\alpha/dL \to 0$ implying $\theta_{\rm tilt} \to \infty$ by Eqs.~(\ref{E:Ddef}) and
(\ref{E:thetatilt}).} \label{F:AC}
\end{figure*}

In this section, we investigate how often BBHs encounter nutational resonances as they inspiral towards merger from the large
separations at which they form.  As the condition $\alpha = 2\pi n$ for integer $n$ defines a nutational resonance, we begin by
calculating $\alpha$ according to Eq.~(\ref{E:alpha}).  Although the parameter space of all BBHs with given masses, spin
magnitudes, and binary separation is four dimensional (corresponding to the two BBH spin directions), two of these dimensions
can be specified by a global rotation of the system about $\mathbf{J}$ and the precessional phase, neither of which affect
$\alpha$ which varies on the radiation-reaction timescale.  For these BBHs (for which $L$ is fixed), $\alpha$ is purely function of
$J$ and $\xi$ for allowed values of these parameters.  We show a contour plot of $\alpha$ for these allowed values in
Fig.~\ref{F:AC}, where the contour lines $\alpha = 2\pi n$ identify nutational resonances.  The largest allowed value of the
magnitude of the total angular momentum $J$ is $J_{\rm max} = L + S_1 + S_2$ and occurs for the ``up-up'' configuration in which
both spins $\mathbf{S}_1$ and $\mathbf{S}_2$ are aligned with the orbital angular momentum $\mathbf{L}$.  Since
$L > S_1 + S_2$ for these BBH masses and spins, the smallest allowed value of $J$ is $J_{\rm min} = L - S_1 - S_2$ and occurs
for the ``down-down'' configuration in which $\mathbf{S}_1$ and $\mathbf{S}_2$ are anti-aligned with $\mathbf{L}$.  The
boundaries of the allowed region in the $J-\xi$ plane are defined by two paths connecting the ``up-up'' and ``down-down''
configurations.  The first of these paths, $\xi_{\rm max}(J)$, connects the maxima of the effective potential $\xi_+(S)$ given by
Eq.~(\ref{E:effpot}).  This path includes the ``down-up'' configuration in which the spin $\mathbf{S}_1$ of the more massive black
hole is anti-aligned with $\mathbf{L}$ while the spin $\mathbf{S}_2$ of the less massive black hole is aligned.  The second path
$\xi_{\rm min}(J)$ connects the minima of the effective potential $\xi_-(S)$.  The allowed region in Fig.~\ref{F:AC} consists of
those BBHs for which $J_{\rm min} \leq J \leq J_{\rm max}$ and $\xi_{\rm min}(J) \leq \xi \leq \xi_{\rm max}(J)$. 

The $n = 1$ and $n = 2$ contours in Fig.~\ref{F:AC} connect points on the $\xi_{\rm min}(J)$ and $\xi_{\rm max}(J)$ curves that
constitute the boundaries of the allowed region.  Because these boundaries correspond to extrema of the effective potential
$\xi_\pm(S)$ (what Schnittman \cite{2004PhRvD..70l4020S} described as spin-orbit resonances), $S$ does not oscillate,
$\Omega_z(S)$ given by Eq.~(\ref{E:Omega_z}) is a constant on the precession timescale, 
and the coefficients $\theta_{Ln}$ given by Eq.~(\ref{E:tLn}) vanish for $n \neq 0$.  The tilt angle $\theta_{\rm tilt}$ given by 
Eq.~(\ref{E:thetatilt}) is proportional to $\theta_{Ln}$ and thus must similarly vanish for $n \neq 0$.  The $n = 1$ and $n = 2$
contours in Fig.~\ref{F:AC} are monotonic functions of both $J$ and $\xi$, so either of these quantities can be used to
parametrize the curves.  We show $\theta_{\rm tilt}(J)$ and $\theta_{\rm tilt}(\xi)$ in the top and right panels of Fig.~\ref{F:AC}.
As expected, $\theta_{\rm tilt}$ vanishes at the endpoints of these curves (the Schnittman spin-orbit resonances) for both nutational
resonances.  The curves $\theta_{\rm tilt}(J)$ and $\theta_{\rm tilt}(\xi)$ are smooth functions for the $n = 1$ resonance, reaching a
maximum $\theta_{\rm tilt} \simeq 7 \times 10^{-4}$ somewhere in the interior of the allowed region.  The corresponding curves for
the $n = 2$ resonance show two sharp spikes where the tilt angle appears to diverge.  These spikes are artifacts of the
approximations used in Section~\ref{S:resonances} and occur where $d\alpha/dL$ and thus $D$ given by Eq.~(\ref{E:Ddef})
vanish.  Since $D$ appears in the denominator of Eq.~(\ref{E:thetatilt}) for $\theta_{\rm tilt}$, the tilt angle correspondingly
diverges.  Physically, points for which both $\alpha = 2\pi n$ and $d\alpha/dL = 0$ correspond to BBHs that are in nutational
resonances and remain in these resonances as they inspiral on the radiation-reaction timescale.  In practice, the quadratic term
in the Taylor expansion of Eq.~(\ref{E:freqT}) will be non-vanishing for these BBHs, implying that the phase $\psi_n$ given by
Eq.~(\ref{E:psiT}) will by cubic rather than quadratic in $t$.  An order-of-magnitude analysis for these BBHs suggests that
$\theta_{\rm tilt}$ will be proportional to $(r/M)^{-1}$ rather than $(r/M)^{-5/4}$ as in Eq.~(\ref{E:thetatilt}), implying somewhat
larger but still finite tilts.

The contours for the $n > 2$ resonances in Fig.~\ref{F:AC} exhibit more complicated behavior.  The $n = 3$ contour begins on
the $\xi_{\rm min}(J)$ boundary, then curves up and to the right until it encounters the solid black curve connecting the ``up-up''
and ``up-down'' configurations identifying those BBHs for which the total spin $\mathbf{S}$ and orbital angular momentum
$\mathbf{L}$ are aligned at $S = S_-$.  For these BBHs, the total angular momentum $\mathbf{J} = \mathbf{L} + \mathbf{S}$ is
also aligned with $\mathbf{L}$ implying that $\alpha$ is undefined as was previously discussed in Section~\ref{S:expansion}
\cite{2016PhRvD..93d4071T}.  Crossing this solid black curve causes $\alpha$ to change discontinuously by $2\pi$, transforming our
$n = 3$ contour into an $n = 4$ contour, another nutational resonance.  For these BBH masses and spins, the ``up-down'' configuration
defining one endpoint of the $\alpha$ discontinuity curve lies in the interior rather than on the $\xi_{\rm min}(J)$ boundary of the
allowed region in the $J-\xi$ plane.  This occurs for binary separations $r_{\rm ud-} < r < r_{\rm ud+}$, where the limits
\begin{equation} \label{E:UDlim}
r_{\rm ud\pm} = \frac{(\sqrt{\chi_1} \pm \sqrt{q\chi_2})^4}{(1 - q)^2} M
\end{equation}
define the range for which the ``up-down'' configuration is unstable to precession to large spin misalignments
\cite{2015PhRvL.115n1102G}.  For these unstable ``up-down'' configurations, the nutation period $\tau$ is infinite, just as it will
take an infinite amount of time for a particle moving in a one-dimensional potential to reach a local maximum (unstable
equilibrium point) which it has just enough energy to access.  Since the precession frequency $\Omega$ remains finite as one
approaches the unstable ``up-down'' configuration while the nutation period $\tau$ diverges, the precession angle
$\alpha = \Omega\tau$ also becomes infinite.  This implies that as one approaches the point in the $J-\xi$ plane corresponding
to the unstable ``up-down'' configuration, one will encounter nutational resonances $\alpha = 2\pi n$ for arbitrarily large values
of $n$.  This is what we see in the inset to the central panel of Fig.~\ref{F:AC}: the contour lines corresponding to nutational
resonances spiral inwards towards the ``up-down'' configuration, with $n$ increasing by an integer each time the solid line
marking the $\alpha$ discontinuity is crossed.  Although $n$ diverges along this spiral, the tilt angle $\theta_{\rm tilt}$ approaches
zero because the BBH spends an increasing fraction of the nutation period with $\mathbf{L}$ closely aligned with $\mathbf{J}$,
implying little tilt in $\langle \mathbf{J}\rangle $ for radiation reaction described by the quadrupole formula of Eq.~(\ref{E:dvecJdt}).

\begin{figure}
\includegraphics[width=1.0\columnwidth]{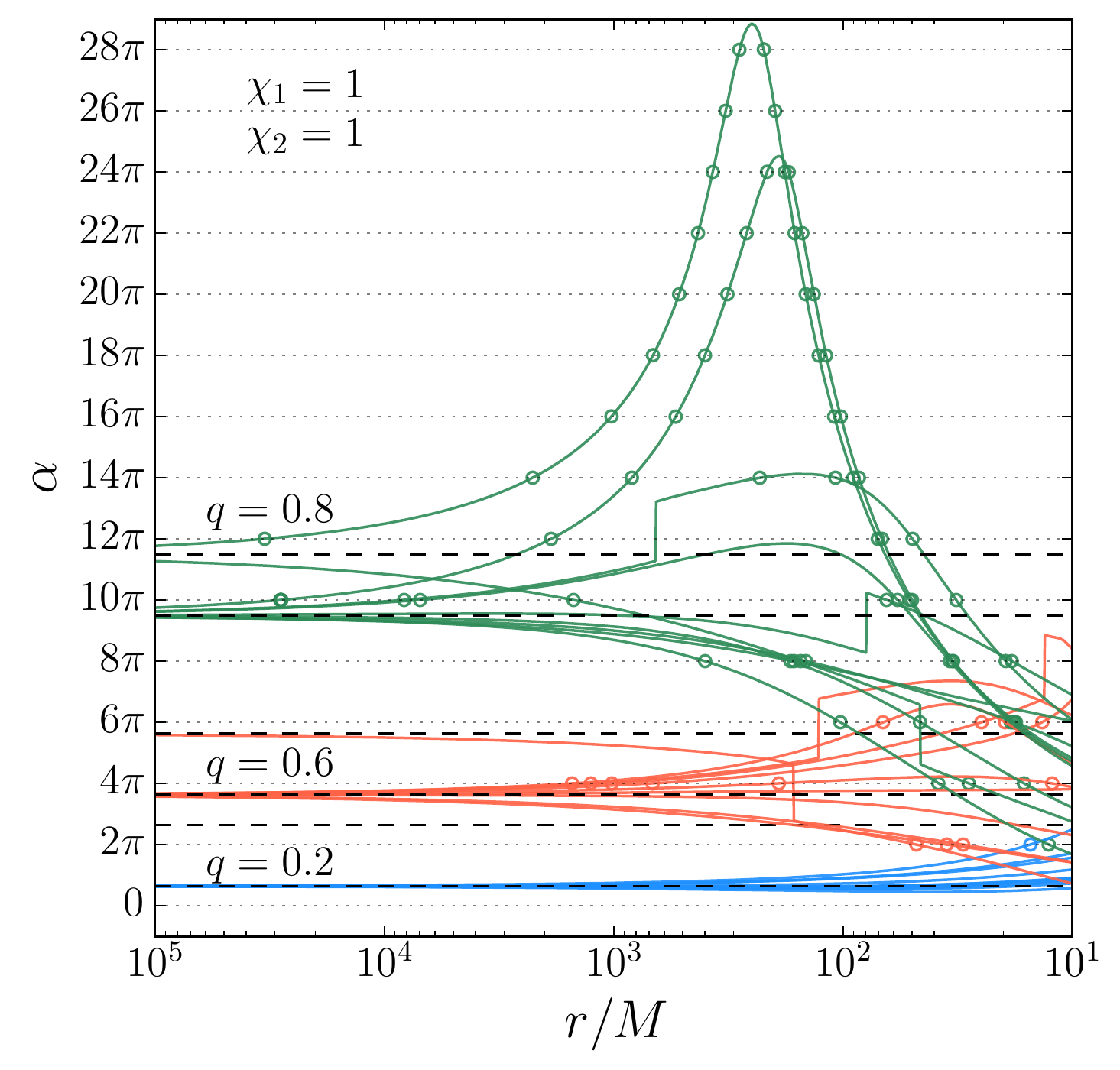}
\caption{The angle $\alpha$ by which the orbital angular momentum $\mathbf{L}$ and total angular momentum $\mathbf{J}$
precess about the $z$ axis during a nutation period $\tau$ as a function of binary separation $r$ for 30 different binaries with
randomly oriented maximal spins ($\chi_i = 1$).  The 10 green, red, and blue curves correspond to BBHs with mass ratios
$q = 0.8$, 0.6, and 0.2 respectively.  The dashed black lines show the asymptotic values $\alpha_{\infty\pm}$ as $r \to \infty$ for
the three mass ratios.  The dotted black lines show the resonance condition $\alpha = 2\pi n$, while the colored circles indicate
separations where the BBHs encounter nutational resonances.} \label{F:30insp}
\end{figure}

Now that we understand which BBHs are in nutational resonances at a given binary separation (for example, $r = 20 M$ in
Fig.~\ref{F:AC}), we can examine when BBHs encounter these resonances as their separation decreases as they inspiral
towards merger.  In Fig.~\ref{F:30insp}, we show $\alpha(r)$ for 30 BBHs (10 each for mass ratios $q = 0.8$, 0.6, and 0.2) with
randomly oriented maximal spins as they inspiral from $r = 10^5 M$ to a final separation $r = 10 M$.  At large separations, we see
that the precession angles $\alpha$ asymptote to one of two different values for each of the three mass ratios; these asymptotic
values are shown by the dashed black lines in Fig.~\ref{F:30insp}.  This surprising result can be understood by recognizing that
the lower PN order spin-orbit coupling dominates over the high-order spin-spin coupling in the limit $r \to \infty$.  In this limit,
the angles between the orbital angular momentum $\mathbf{L}$ and the BBH spins $\mathbf{S}_1$ and $\mathbf{S}_2$ are fixed
to their asymptotic values $\theta_{1\infty}$ and $\theta_{2\infty}$, $\mathbf{L}$ and the total angular momentum $\mathbf{J}$ are
both nearly aligned with the $z$ axis, and the two spins precess about this axis with respective frequencies
\cite{1994PhRvD..49.6274A,1995PhRvD..52..821K}
\begin{subequations} \label{E:SpinOrb}
\begin{align}	
\Omega_1 &= \frac{(4+3q)\eta}{2M} \left( \frac{r}{M} \right)^{-5/2}, \label{E:SO1} \\
\Omega_2 &= \frac{(4+3/q)\eta}{2M} \left( \frac{r}{M} \right)^{-5/2}. \label{E:SO2}
\end{align}
\end{subequations}
Unless the BBHs masses are precisely equal, the mass ratio $q < 1$ and the spin of the less massive black hole precesses
faster ($\Omega_2 > \Omega_1$).  If the components of $\mathbf{S}_1$ and $\mathbf{S}_2$ perpendicular to the $z$ axis are
aligned at $t = 0$, they will first realign (return to their initial relative orientations) after a nutation period $\tau$.  Over this interval,
the faster spin $\mathbf{S}_2$ will precess about $z$ by an additional $2\pi$ radians compared to the slower spin
$\mathbf{S}_1$:
\begin{equation} \label{E:TauCon}
(\Omega_2-\Omega_1)\tau = 2\pi~.
\end{equation}
In order for $\mathbf{J}$ to remain nearly aligned with the $z$ axis
($\theta_J \ll \theta_L$), $\mathbf{L}$ must have a component $\mathbf{L}_\perp$ in the $xy$ plane anti-aligned with the
component $\mathbf{S}_\perp$ of the total spin in this plane.  If
$S_{1\perp} = S_1\sin\theta_{1\infty} > S_{2\perp} = S_2\sin\theta_{2\infty}$, $\mathbf{S}$ and thus $\mathbf{L}$
will precess about the $z$ axis over the nutation period $\tau$ by an angle
\begin{equation} \label{E:alpinf-}
\alpha_{\infty-} = \Omega_1\tau = \frac{2\pi\Omega_1}{\Omega_2-\Omega_1} = \frac{2\pi q(4 + 3q)}{3(1 - q^2)}~,
\end{equation}
which we have derived using Eqs.~(\ref{E:SpinOrb}) and (\ref{E:TauCon}).  If $S_1\sin\theta_{1\infty} < S_2\sin\theta_{2\infty}$, the
asymptotic precession angle will instead be given by
\begin{equation} \label{E:alpinf+}
\alpha_{\infty+} = \Omega_2\tau = \alpha_{\infty-} + 2\pi = \frac{2\pi(4q + 3)}{3(1 - q^2)}~.
\end{equation}
If the BBHs have isotropically oriented spins with magnitudes $S_1$ and $S_2$, the fraction of binaries
for which $\alpha$ asymptotes to $\alpha_{\infty+}$ for $S_1 > S_2$ is
\begin{equation} \label{E:F+G}
f_+ = \frac{|S_1^2 - S_2^2|}{4S_1S_2} [\sinh (2\cosh^{-1}C) - 2\cosh^{-1}C]~,
\end{equation}
while for $S_1 < S_2$ it is
\begin{equation} \label{E:F+L}
f_+ = \frac{|S_1^2 - S_2^2|}{4S_1S_2} [\sinh (2\sinh^{-1}C) + 2\sinh^{-1}C]~,
\end{equation}
where in both expressions $C \equiv S_1/|S_1^2 - S_2^2|^{1/2}$.  For the three mass ratios $q = \{ 0.8, 0.6, 0.2\}$ in
Fig.~\ref{F:30insp}, Eqs.~(\ref{E:alpinf-}) through (\ref{E:F+G}) imply $\alpha_{\infty-} = \{ 9.48\pi, 3.63\pi, 0.64\pi\}$,
$\alpha_{\infty+} = \{ 11.48\pi, 5.63\pi, 2.64\pi\}$, and $f_+ = \{ 0.15, 0.0444, 5.34 \times 10^{-4}\}$.  These values are consistent
with the horizontal dashed lines in Fig.~\ref{F:30insp} and that $\{ 2/10, 1/10, 0/10 \}$ of the binaries asymptote to
$\alpha_{\infty+}$ for $q = \{ 0.8, 0.6, 0.2\}$.

As the BBHs in Fig.~\ref{F:30insp} inspiral from large separations towards merger, they encounter nutational resonances marked
by small colored circles whenever $\alpha = 2\pi n$.  BBHs with mass ratios for which $\alpha_{\infty\pm}$ is close to an integer
multiple of $2\pi$ are most likely to encounter nutational resonances at large binary separations.  We also see several
discontinuous jumps in $\alpha$ by $\pm2\pi$ corresponding to configurations in which the orbital angular momentum
$\mathbf{L}$ and total angular momentum $\mathbf{J}$ are either aligned or anti-aligned at $S = S_+$ or $S_-$.  According to
Eq.~(\ref{E:UDlim}), the BBHs with mass ratios $q = \{ 0.8, 0.6, 0.2\}$ in Fig.~\ref{F:30insp} enter the regime where the ``up-down''
configuration is unstable for binary separations $r < r_{\rm ud+} \simeq \{ 322 M, 62 M, 6.85 M \}$.  The large peak values
$\alpha > 24\pi$ occurring at $r \lesssim r_{\rm ud+}$ for two of the $q = 0.8$ binaries in Fig.~\ref{F:30insp} result from close
approaches in the $J-\xi$ plane to the unstable ``up-down'' configuration for which $\alpha \to \infty$.  The key point to take away
from Fig.~\ref{F:30insp} is that most binaries with large spins and $q \gtrsim 0.6$ encounter one or more nutational resonances
during their inspiral, and many of these resonances occur at $r < 100 M$ where tilts are comparatively large and GWs for
solar-mass BBHs are emitted at frequency detectable by ground-based GW observatories like LIGO.

\begin{figure*}\centering
\includegraphics[width=1.7\columnwidth]{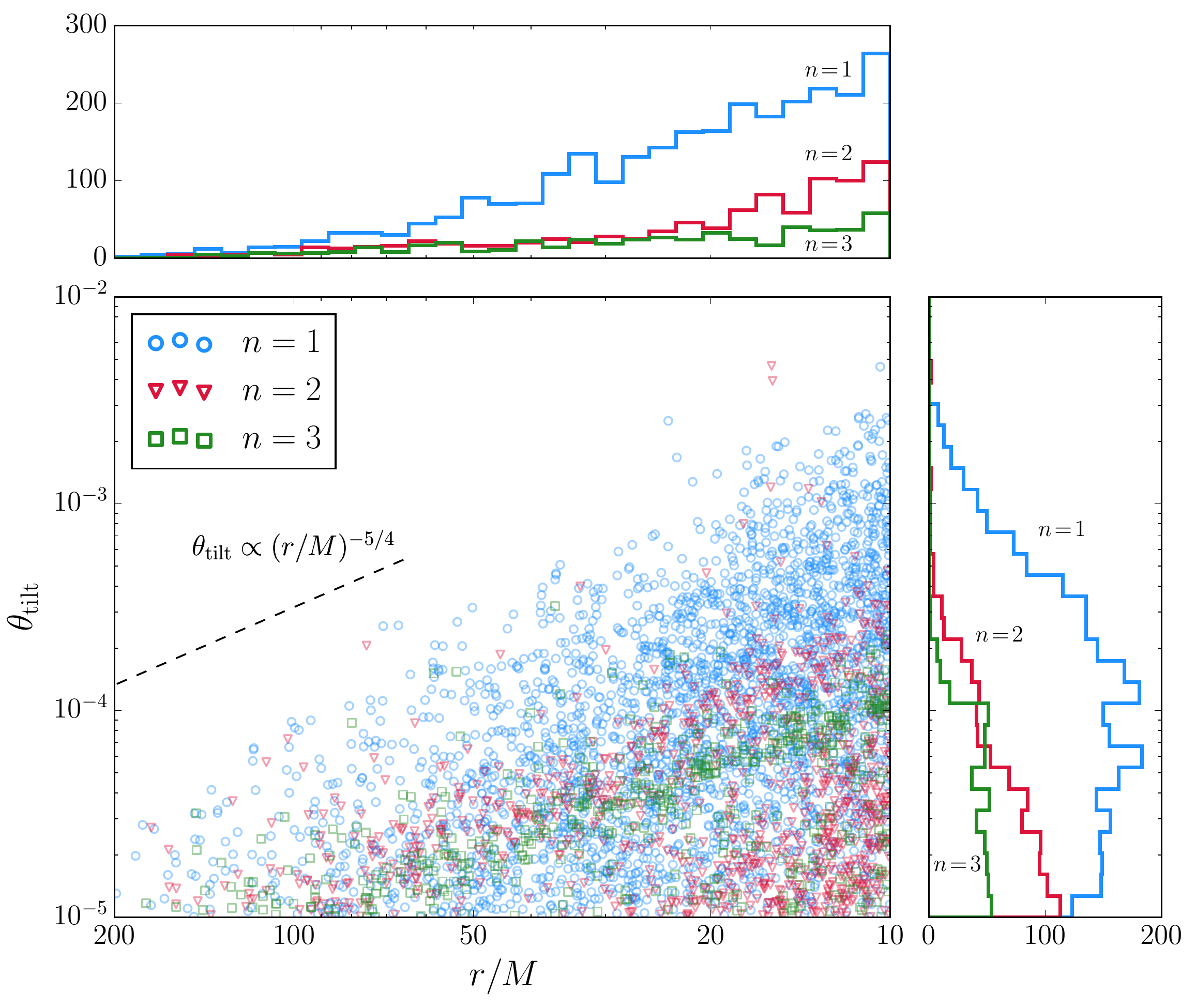}
\caption{Distribution of nutational resonances as a function of binary separation $r$ and tilt angle $\theta_{\rm tilt}$ for
$5 \times 10^4$ binaries with isotropically oriented spins and flat distributions of mass ratios and dimensionless spins in the
ranges $0.1 < q < 1$ and $0.1 < \chi_i < 1$.  The central panel is a scatter plot in which each blue circle, red triangle, and green
square corresponds to a $n = \{ 1, 2, 3 \}$ nutational resonance encountered by one of the BBHs.  The dashed black line
$\theta_{\rm tilt} \propto (r/M)^{-5/4}$ shows that the scaling of the largest tilt angles with binary separation agrees with the analytic
prediction of Eq.~(\ref{E:thetatilt}).  The top and right panels show histograms generated by binning this scatter plot
of the nutational resonances as functions of binary separation $r$ and tilt angle $\theta_{\rm tilt}$ for each value of $n$.  The
binaries encounter a total of $N = \{ 2717, 923, 517 \}$ resonances for $n = \{ 1, 2, 3 \}$ with $\theta_{\rm tilt} > 10^{-5}$ in the
range $200 M > r > 10 M $, implying that $\sim8.3\%$ of the BBHs encounter such resonances.} \label{F:hist}
\end{figure*}

Having examined how $\alpha$ evolves with binary separation for the 30 binaries show in Fig.~\ref{F:30insp}, we now broaden
our sample to $5 \times 10^4$ binaries with a flat distribution of mass ratios in the range $0.1 < q < 1$ and isotropic spins with
a flat distribution of dimensionless magnitudes in the range $0.1 < \chi_i < 1$.  In Fig.~\ref{F:hist}, we show all of the nutational
resonances with $|n| \leq 3$ and $\theta_{\rm tilt} > 10^{-5}$ encountered by these binaries as they inspiral from $r = 200 M$ to
$r = 10 M$.  No resonances with $n \leq 0$ were observed, suggesting that such resonances may not exist although we have not
found a mathematical proof of their non-existence.  A total of 4157 nutational resonances were found during these
$5 \times 10^4$ inspirals (an incidence of 8.3\%), with most occurring at $r \lesssim 50 M$ as shown by the histogram in the top
panel of Fig.~\ref{F:hist}.  The previous sample shown in Fig.~\ref{F:30insp} suggests that BBHs with comparable masses should
account for the majority of these nutational resonances because the steeper slopes of their $\alpha(r)$ curves should increase
the probability that they cross an $\alpha = 2\pi n$ line signaling a nutational resonance.  There are 2717 $n = 1$ resonances
(65.4\% of the total) with a broad range of tilts, including a tail extending to $\theta_{\rm tilt} > 10^{-3}$ for $r \lesssim 20 M$ as
shown by the histogram in the right panel of Fig.~\ref{F:hist}.  The largest tilt angles appears to scale with binary
separation as $\theta_{\rm tilt} \propto (r/M)^{-5/4}$ consistent with the analytic estimate of Eq.~(\ref{E:thetatilt}).  The 923 $n = 2$
and 517 $n = 3$ resonances constitute smaller fractions of the total (22.2\% and 12.4\% respectively) and generally lead to
smaller tilts $\theta_{\rm tilt} < 10^{-3}$.  Although there may be finely tuned resonances missing from our sample with even
larger tilts (such as those with $d\alpha/dL = 0$ indicated by the spikes in the top and right panels of Fig.~\ref{F:AC}), the results
shown in Fig.~\ref{F:hist} suggest that tilts from exact resonances at binary separations $r > 10 M$ are too small to have
significant astrophysical consequences or detectable GW signatures.  However, we will show in the next section that the large
tilts associated with transitional precession \cite{1994PhRvD..49.6274A} can be interpreted as a consequence of an approximate
$n = 0$ nutational resonance.

\section{Transitional precession as an approximate nutational resonance} \label{S:trans}
	
\begin{figure*}[p]
\includegraphics[width=2\columnwidth]{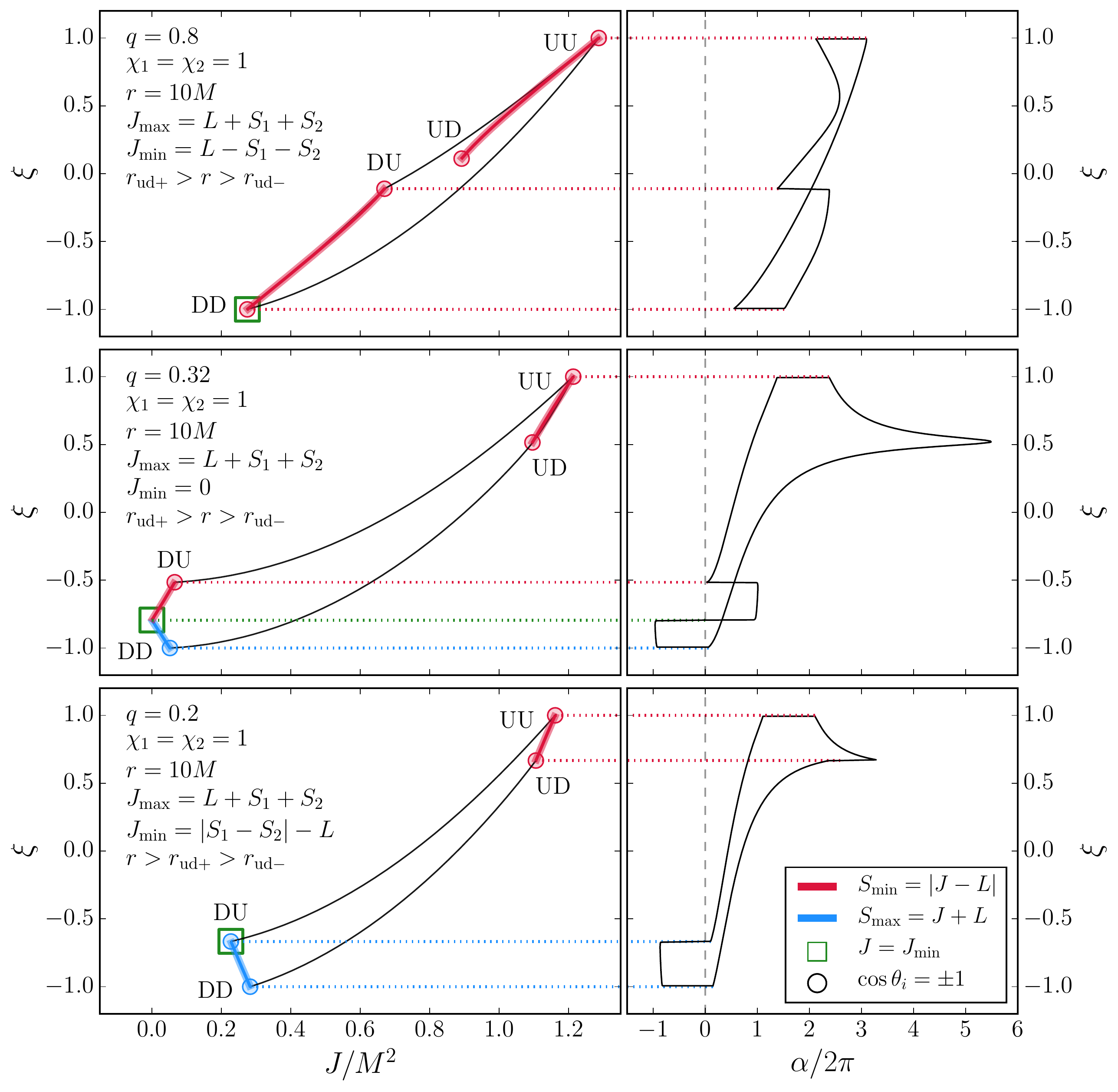}
\caption{{\it Left panels:} Boundaries of the allowed region in the $J-\xi$ plane for BBHs with binary separations of $r = 10 M$,
maximal spin magnitudes, and mass ratios of $q = 0.8$ (top panel), 0.32 (middle panel), and 0.2 (bottom panel). The BBHs along
the red curves have the orbital angular momentum $\mathbf{L}$ aligned with the total angular momentum $\mathbf{J}$ once per
nutation period, when $S = S_- = S_{\rm min} = |J - L|$, while the BBHs along the blue curves have $\mathbf{L}$ aligned with
$\mathbf{J}$ once per nutation period when $S = S_+ = S_{\rm max} = J + L$.  The four circles in each panel indicate the BBHs
for which the spins $\mathbf{S}_i$ are both either aligned or anti-aligned with $\mathbf{L}$; the circles labeled ``UU'', ``UD'', ``DU'', and
``DD'' correspond to the up-up, up-down, down-up, and down-down configurations respectively.  The green squares indicate the
minimum allowed value of $J$ for each mass ratio.  {\it Right panels:} The precession angle $\alpha$ over a nutation period
$\tau$ as a function of the projected effective spin $\xi$ along the boundaries of the allowed regions.  The horizontal dotted lines
indicate values of $\xi$ at which the red and blue curves intersect the boundaries, while the vertical dashed line indicates
$\alpha = 0$.} \label{F:jumps}
\end{figure*}

The tilt angles $\theta_{\rm tilt}$ shown in Fig.~\ref{F:hist} are disappointingly small if we ever hope to measure their observational
consequences.  The coefficients $\theta_{Ln}$ shown in Fig.~\ref{F:coeff} are several orders of magnitude larger for $n = 0$ and
$n = -1$ than the other coefficients, suggesting from Eq.~(\ref{E:thetatilt}) that the tilt angles at $n = 0$ or $n = -1$ nutational
resonances would be similarly larger, perhaps even of order unity, if such resonances could be found.  Our investigation of the
$\alpha = 2\pi n$ contours in Fig.~\ref{F:AC} suggests that if $n = 0$ or $n = -1$ contours exist, they will intersect the boundaries
of the allowed region in the $J-\xi$ plane.  To test this possibility, we plot these boundaries and the value of $\alpha$ along them
for BBHs with maximal spins, binary separations of $r = 10 M$, and three different mass ratios $q = \{ 0.8, 0.32, 0.2 \}$ in
Fig.~\ref{F:jumps}.  These three mass ratios provide examples of the three alternative values of $J_{\rm min}$, the minimum
allowed magnitude of the total angular momentum \cite{2015PhRvD..92f4016G}.  If $L > S_1 + S_2$, the minimum allowed
magnitude of $\mathbf{J} = \mathbf{L} + \mathbf{S}_1 + \mathbf{S}_2$ is $L - S_1 - S_2$ as is the case for $q = 0.8$,
$\chi_1 = \chi_2 = 1$, and $r = 10 M$ as seen in the top panel of Fig.~\ref{F:jumps}.  This value of $J_{\rm min}$, indicated by the
green square, corresponds to the ``down-down'' configuration indicated by one of the four circles showing the four
configurations in which the BBH spins $\mathbf{S}_i$ are both either aligned or anti-aligned with $\mathbf{L}$.  The right side of
this panel shows $\alpha(\xi)$ as we circulate around the boundary of the allowed region in the $J-\xi$ plane.  The continuous
curve connecting $\xi = \pm1$ corresponds to the right edge of the allowed region, while the other two curves correspond to the
left edge of the allowed region.  At each of the three circles on the boundary of the allowed region (the ``up-up'', ``down-up'', and
``down-down'' configurations), the value of $\alpha$ changes discontinuously by $\pm2\pi$ because of the coordinate
discontinuity discussed previously.  As $r_{\rm ud-} < r < r_{\rm ud+}$ according to Eq.~(\ref{E:UDlim}) for this choice of
parameters, the unstable ``up-down'' configuration lies in the interior of the allowed region and thus does not lead to a
discontinuity in $\alpha$ along the boundary.  It is important to note that the red curve connecting the ``down-up'' and
``down-down'' configurations denoting BBHs for which $\alpha$ is undefined lies in the interior of the allowed region, although it
is so close to boundary as to appear indistinguishable from it in this figure.  The red and blue curves in the middle and bottom
panels are also in the interior of the allowed region despite their close proximity to the boundary.

We now examine the middle panel of Fig.~\ref{F:jumps} which differs from the top panel because the mass ratio has been
reduced to $q = 0.32$.  For this mass ratio, $|S_1 - S_2| < L < S_1 + S_2$ implying that the three vectors $\mathbf{L}$,
$\mathbf{S}_1$, and $\mathbf{S}_2$ can form the sides of a triangle and thus their sum $\mathbf{J}$ can vanish.  This is 
equivalent to the statement $J_{\rm min} = 0$ as indicated by the green square in the middle panel.  The precession angle
$\alpha$ is undefined for $\mathbf{J} = 0$ since the $z$ axis cannot be defined to point in the direction of the
precession-averaged value of a vanishing quantity.  This $J = 0$ configuration is connected to the ``down-up'' and ``down-down''
configurations (for which $\alpha$ is also undefined) by red and blue curves indicating BBHs for which the total spin
$\mathbf{S}$ is aligned or anti-aligned with $\mathbf{L}$ at $S = S_-$ or $S = S_+$ respectively.  Let us now consider the right
side of the middle panel in which we again show $\alpha(\xi)$ as we circulate around the boundary of the allowed region in the
$J-\xi$ plane.  As in the top panel, the single continuous curve connecting $\xi = \pm1$ corresponds to the right edge of the
allowed region.  Because $r_{\rm ud+} \simeq 13 M$ for this mass ratio is barely above the binary separation $r = 10 M$, the
unstable ``up-down'' configuration is very close to the right edge of the allowed region and $\alpha$ gets very large near this
configuration as was previously seen in Figs.~\ref{F:AC} and \ref{F:30insp}.  The three other discontinuous curves $\alpha(\xi)$
correspond to the three pieces of the left edge of the allowed region: the first piece connects the ``down-down'' and $J = 0$
configurations, the second piece connects the $J = 0$ and ``down-up'' configurations, and the long third piece of the left boundary
connects the ``down-up'' and ``up-up'' configurations.  As in the top panel, we see that $\alpha$ experiences discontinuous jumps
by $\pm2\pi$ at the ``up-up'', ``down-up'', and ``down-down'' configurations.  However, as we trace along the left edge of the
allowed region and pass through the $J = 0$ configuration, we cross both the red and blue curves leading to a discontinuity by
$\pm4\pi$ in $\alpha$ as seen in the right side of the panel (twice the size of the other discontinuities).  

Although $\alpha$ is undefined for the $J = 0$ configuration, we can consider the value of $\alpha$ in the neighborhood of this
point of the $J-\xi$ plane.  Because the red and blue curves are so close to the left edge of the allowed region, the vast majority
of this neighborhood will lie in between the red and blue curves where $\alpha$ has experienced only half of the $\pm4\pi$
discontinuity that would result from crossing both curves.  Examining the midpoint of the $\alpha$ discontinuity at $J = 0$ on the
right side of the middle panel of Fig.~\ref{F:jumps}, we see that most of the neighborhood of this point has $\alpha \simeq 0$
making it an approximate $n = 0$ nutational resonance.  The large size of the $\theta_{L0}$ coefficient in Fig.~\ref{F:coeff}
compared to those with $n \geq 1$ suggests that this approximate $n = 0$ nutational resonance should lead to a much larger
tilt than those found for the $n = \{1, 2, 3\}$ resonances shown in Fig.~\ref{F:hist}.  In fact, this large tilt is already well known to the
relativity community as the transitional precession described in Apostolatos {\it et al.}~\cite{1994PhRvD..49.6274A}!  Fig.~9 in that
paper shows the evolution of $\mathbf{\hat{L}}$ for a binary with $q = 0.1$, $\mathbf{S}_2 = 0$, and maximal spin
$\mathbf{S}_1$ nearly anti-aligned with $\mathbf{L}$ as the binary inspirals from $r = 330 M $ to $r = 6 M$.  This figure looks
unmistakably like the Euler spirals of Figs.~\ref{F:Jperp_terms} and \ref{F:Euler} of our paper, although Apostolatos {\it et al.} were
not able to obtain our analytic solution of Eq.~(\ref{E:Jperp_res}).  While the Euler spirals at nutational resonances with
$n \geq 1$ can only be identified using our new expansion because they are subdominant to the non-resonant terms with
$n = \{-1, 0\}$, the large tilt resulting from the approximate $n = 0$ resonance during transitional precession can be seen without our
expansion because it involves the dominant term.  We have thus demonstrated that the well-known large tilts during transitional
precession, illustrated for the special case $\mathbf{S}_2 = 0$ in Apostolatos {\it et al.}~\cite{1994PhRvD..49.6274A}, can occur
for $J \simeq 0$ even if $\mathbf{S}_2 \neq 0$.  They are in fact special cases of the more general nutational resonances for
arbitrary $n$ that are far more frequently encountered during generic misaligned inspirals.  The middle panel of
Fig.~\ref{F:jumps} suggests that most of the neighborhoods of the ``down-up'' and ``up-up'' configurations should similarly be
approximate $n = 0$ resonances.  However, the near alignment or anti-alignment of both BBH spins with $\mathbf{L}$ in these
neighborhoods may lead to small values of the $\theta_{L0}$ coefficients and corresponding tilts.  Future investigations
searching for large tilts in $\langle \mathbf{J} \rangle$ should consider configurations that are approximate $n = \{ -1, 0 \}$ nutational
resonances.

For completeness, we show the third possibility for $J_{\rm min}$ in the bottom panel of Fig.~\ref{F:jumps}.  For these binaries
with the even smaller mass ratio $q = 0.2$, $S_1 > L + S_2$ and therefore $J_{\rm min} = S_1 - S_2 - L > 0$.  The
$J = J_{\rm min}$ configuration coincides with the ``down-up'' configuration in this case as seen by the overlapping circle and
green square.  For this mass ratio and binary separation, $r > r_{\rm ud\pm}$ implying that the ``up-down'' configuration is stable
and lies on the right edge of the allowed region in the $J-\xi$ plane.  Examining $\alpha$ along the boundary of the allowed
region as shown on the right side of the bottom panel, we see that there is no longer a continuous curve $\alpha(\xi)$ connecting
$\xi = \pm1$ because of the new $\alpha$ discontinuity on the right edge of the allowed region associated with the now stable
``up-down'' configuration. The left and right edges of the allowed region are each associated with two discontinuous $\alpha(\xi)$
curves, with four jumps in $\alpha$ by $\pm2\pi$ corresponding to the four configurations with $\cos\theta_i = \pm1$ as shown
by the horizontal dotted lines.  Nature seems to have frustrated our efforts to discover exact $n = 0$ or $n = -1$ nutational
resonances; the $\alpha$ discontinuities jump across $\alpha = 0$, while $\alpha$ never gets quite negative enough for an exact
$n = -1$ resonance.  Although the results shown in Figs.~\ref{F:hist} and \ref{F:jumps} suggest that such resonances do not exist,
we have not been able to derive a mathematical proof to this effect.

\section{Discussion} \label{S:disc}

This paper seeks to provide qualitative and quantitative insight into the evolution of the orbital angular momentum $\mathbf{L}$
and total angular momentum $\mathbf{J}$ in the PN regime for generic BBHs (unequal masses, two misaligned spins).  We rely
extensively on our earlier work \cite{2015PhRvL.114h1103K,2015PhRvD..92f4016G} in which we derived analytic solutions
to the 2PN spin-precession equations for generic binaries in the absence of radiation reaction.  These solutions showed that the
relative orientations of $\mathbf{L}$ and the BBH spins $\mathbf{S}_i$ could be fully specified by a single degree of freedom and
that the magnitude $S$ of the total spin $\mathbf{S} = \mathbf{S}_1 + \mathbf{S}_2$ was a useful coordinate for describing this
degree of freedom.  In the absence of radiation reaction, $\mathbf{J}$ is fixed (and thus equal to its precession-averaged value
$\langle \mathbf{J} \rangle$) and can be used to define the $z$ axis in an inertial reference frame.  Without loss of generality, we
can choose $x$ and $y$ axes in the plane perpendicular to $\mathbf{\hat{z}}$.  The direction of $\mathbf{L}$ in this frame can be
specified by the spherical coordinates $\theta_L$ and $\Phi_L$.  For generic BBHs at 2PN order, $\mathbf{L}$ will both precess
(evolution of $\Phi_L$) and nutate (evolution of $\theta_L$).  Over a nutation period $\tau$, $S$ will oscillate back and forth
between its extrema $S_\pm$ set by the effective potential $\xi_\pm(S)$, and $\theta_L$ will similarly oscillate according to
Eq.~(\ref{E:costhetaL}).  While it nutates, $\mathbf{L}$ will precess at the time-dependent precession frequency
$d\Phi_L/dt = \Omega_z(S)$ given by Eq.~(\ref{E:Omega_z}), precessing by a total angle $\alpha$ over a full nutation period
$\tau$.

In this paper, we used $\tau$ and $\alpha$ to define the nutation frequency $\omega \equiv 2\pi/\tau$ and average precession
frequency $\Omega \equiv \alpha/\tau$ that characterize the evolution of $\mathbf{L}$ on the precession timescale.  We derived
Eq.~(\ref{E:inbas}), a new series expansion for the component of $\mathbf{L}$ in the $xy$ plane, in which each term is a vector of
length $|\theta_{Ln}|$ that precesses about the $z$ axis with frequency $\Omega - n\omega$.  Fig.~\ref{F:comp} demonstrates
that just the two dominant terms in this series can very accurately describe the evolution of $\mathbf{L}$ even when it exhibits
seemingly complicated precession and nutation.

Radiation reaction modifies the above analysis because the total angular momentum $\mathbf{J}$ no longer remains constant.
However, with the $z$ axis defined to point in the direction of the precession-averaged total angular momentum
$\langle \mathbf{J} \rangle$, our new series expansion for $\mathbf{L}$ remains approximately valid because the angle
$\theta_J$ between the instantaneous $\mathbf{J}$ and its precession average $\langle \mathbf{J} \rangle$ is suppressed
compared to $\theta_L$ by the ratio of the precession and radiation-reaction timescales
$t_{\rm pre}/t_{\rm RR} \propto (r/M)^{-3/2} \ll 1$.  This allows us to approximately equate the angles between $\mathbf{L}$ and the
two vectors $\mathbf{J}$ and $\langle \mathbf{J} \rangle$ ($\theta_{Lz} \simeq \theta_L$ for the angles depicted in
Fig.~\ref{F:frame}).  To use our series expansion with non-vanishing radiation reaction, we need only allow the coefficients
$\theta_{Ln}$ and frequencies $\omega$ and $\Omega$ to vary on the radiation-reaction timescale, replacing the phases of
each term in the expansion by the time integrals of the now varying frequencies.  The excellent agreement between our expansion and
direct numerical integration of the spin-precession equations shown in Fig.~\ref{F:comp} was obtained with just this prescription for
radiation reaction.  Because the coefficients and frequencies of our expansion only vary on the radiation-reaction timescale
$t_{\rm RR}$ (unlike $\mathbf{L}$ itself which evolves on the precession timescale $t_{\rm pre}$), our expansion may provide vast
computation savings if $\mathbf{L}(t)$ needs to be calculated over an entire inspiral to generate gravitational waveforms.

The proportionality between $d\mathbf{J}/dt$ and $\mathbf{L}$ for radiation reaction described by the quadrupole formula
(which also holds for the 1PN corrections to this formula \cite{1995PhRvD..52..821K}) implies that our new series expansion
for $\mathbf{L}$ can also be used to describe $d\mathbf{J}/dt$ as seen in Eq.~(\ref{E:dJperpdt}).  To understand the evolution of
$\mathbf{J}$ on the precession timescale, we need only integrate this expansion while holding the coefficients and
frequencies (which vary on the radiation-reaction timescale) constant.   This analytic integration breaks down whenever
$\Omega - n\omega = 0$ (mathematically equivalent to $\alpha = 2\pi n$), since this combination of frequencies appears in the
denominator of the integral.  We identify this condition as a nutational resonance, since the average precession frequency
$\Omega$ is an integer multiple of the nutation frequency $\omega$.  Physically, this breakdown occurs because if
$\alpha = 2\pi n$, the component of $\mathbf{L}$ in the $xy$ plane will return to its initial value after a nutation period $\tau$.  The
total angular momentum radiated in successive nutation periods will therefore point in the same direction and add constructively,
causing $\langle \mathbf{J} \rangle$ to tilt into the $xy$ plane.  Although an exact nutational resonance
requires a finely tuned value of $\alpha$, the sample of $5 \times 10^4$ BBHs shown in Fig.~\ref{F:hist} shows that $\sim10\%$
of BBHs encounter a nutational resonance with $n = \{ 1, 2, 3 \}$ as they inspiral from $r = 200 M$ to $r = 10 M$.  However, the
tilt angles $\theta_{\rm tilt}$ associated with these resonances are typically less than $10^{-3}$ radians even at small $r$
because the coefficients $\theta_{Ln}$ to which these tilts are proportional are highly subdominant to the non-resonant $n = -1$
and $n = 0$ terms in the series expansion for $\mathbf{J}$.

Although we have not found any exact $n = 0$ nutational resonances, a careful examination of BBHs in the neighborhood of the
$J = 0$ configuration as shown in the middle panel of Fig.~\ref{F:jumps} reveals that most of these BBHs are in an approximate
$n = 0$ resonance leading to large tilts.  Our identification of this approximate nutational resonance is in fact just a new
description of the familiar phenomenon of transitional precession identified by
Apostolatos {\it et al.}~\cite{1994PhRvD..49.6274A}.  In that paper, transitional precession was derived in the limit that
$\mathbf{S}_2 = 0$ and $\mathbf{S}_1 \simeq -\mathbf{L}$, but we have shown that it also applies for most configurations where
the total spin $\mathbf{S} \simeq -\mathbf{L}$, even if both BBHs spins are near maximal.  A more systematic investigation of
other mass ratios and spin magnitudes could potentially discover other approximate $n = -1$ and $n = 0$ resonances where
$\langle \mathbf{J} \rangle$ experiences large tilts, a significant source of error in the construction of gravitational waveforms
\cite{2017PhRvD..95j4004C}.

We hope that the insights provided in this paper, particularly our elegant new series expansion for $\mathbf{L}$ in
Eq.~(\ref{E:inbas}), will prove useful for future calculations of GW emission and more general astrophysical studies of BBHs.
Although the precession of $\mathbf{L}$ in an inertial frame has long been recognized for systems with misaligned spins, the
nutation of $\mathbf{L}$ has received less attention.  This nutation, a consequence of multiple non-vanishing terms in
Eq.~(\ref{E:inbas}), will likely generate distinctive observational signatures in both gravitational waveforms and astrophysical
phenomena like the jets and circumbinary disks associated with accreting supermassive BBHs
\cite{2002Sci...297.1310M,2013ApJ...774...43M,2015MNRAS.451.3941G}.  Whether these signatures are large and unambiguous enough to be detected
remains an open question, but one we hope may be addressed by the wealth of observations that will be provided by upcoming GW
and electromagnetic surveys.

\acknowledgements
M. K. is supported by the Alfred P. Sloan Foundation Grant No. FG-2015-65299 and NSF Grant No. PHY-1607031.  D. G. is
supported by NASA through Einstein Postdoctoral Fellowship Grant No. PF6-170152 awarded by the Chandra X-ray Center,
which is operated by the Smithsonian Astrophysical Observatory for NASA under contract NAS8-03060. 
Computations were performed on the Caltech computer cluster ``Wheeler,'' supported by the Sherman Fairchild Foundation and Caltech. Partial support is acknowledged by NSF Award CAREER PHY-1151197.

\bibliography{Resonances}

\end{document}